\documentclass[%
 aip,
 jmp,%
 amsmath,amssymb,
 preprint,%
]{revtex4-1}

\usepackage{graphicx}
\usepackage{dcolumn}
\usepackage{bm}
\usepackage{multirow}

\usepackage{amsmath}
\usepackage{amssymb}
\usepackage{color,soul}
\begin{document}

\title[Towards Hybrid Density Functional Calculations of Molecular Crystals via Fragment-Based Methods]{Towards  Hybrid Density Functional Calculations of Molecular Crystals via Fragment-Based methods}

\author{Oleksandr A. Loboda}
\altaffiliation[On leave from]{:\ A.V. Dumansky Institute of Colloid and Water Chemistry, National Academy of Sciences of Ukraine, 
Vernadsky bld. 42, Kyiv-142, Ukraine }
\author{Grygoriy A. Dolgonos}
\author{A. Daniel Boese}
\email{adrian\_daniel.boese@uni-graz.at}
 \homepage{http://www.chemie.uni-graz.at/en/quantum-chemistry}
\affiliation{ 
Institute of Chemistry, University of Graz, Heinrichstrasse 28/IV, A-8010 Graz, Austria}

\date{\today}

\begin{abstract}
We introduce and employ two QM:QM schemes (a quantum mechanical method embedded into another quantum mechanical method) and report their performance for the X23 set of molecular crystals.
We furthermore present the theory to calculate the stress tensors necessary for the computation of optimized  cell volumes of molecular crystals
and compare all results to those obtained with various density functionals and more approximate methods. Our QM:QM calculations with PBE0:PBE+D3, PBE0:PBE+MBD, and B3LYP:BLYP+D3 yield at a reduced computational cost lattice energy errors close to the ones of the parent hybrid density functional method, whereas for cell volumes, the errors of the QM:QM scheme methods are in between the GGA and hybrid functionals. 

\end{abstract}

\pacs{Valid PACS appear here}
\keywords{Density Functional Theory, Molecular Crystals, Fragment-based Methods, Hybrid Functionals}
\maketitle


\section{\sffamily \Large Introduction}


The computational description of molecular crystals has come a long way in the last two decades:
this can especially be seen when considering its most important application, crystal structure prediction (CSP).
The ultimate goal of the crystal structure prediction is to explore all possible polymorphs, co-crystals, salts,
solvates (hydrates) of several molecules, based solely on the minimal information of its Lewis structure\cite{amreilly,ccdc}.
Only the two-dimensional, schematic diagram of some organic molecules in the gas phase were revealed to the community
(together with some basic information about known polymorphism and the crystallization conditions) with the request for a competitive CSP.
This was actually done in the so-called blind tests which were organised by the Cambridge Crystallographic Data Center (CCDC).
For the first blind test in 2000, "no program gave consistently reliable results"\cite{ccdc} when
crystal structures of rather small and simple molecules up to 28 atoms had to be compared to experiment.
In contrast, at the last blind test in 2015, "All of the targets, apart from a single potentially disordered
Z' = 2 polymorph of the drug candidate, were predicted by at least one submission."\cite{amreilly}
In this test, the molecules had a considerable larger complexity than for previous blind tests:
Flexible molecules with more than 60 atoms were predicted, together with a polymorph, a salt, and a co-crystal.
There are several aspects which lead to this remarkable success, which are a) the improvement of the description of the monomers by better
ab initio methods b) construction of better, or even automatic force fields for which the searches are performed c) the development of
 enhanced search algorithms by itself and finally d) refinement methods, which further optimize the top ranked structures obtained by the force fields
 in step b). We will concentrate on improving the last step d), which is to develop electronic structure methods for the computation
 of molecular crystals.
 
Currently, density functional theory including dispersion interactions is the method of choice when performing step d) of the CSP.
More approximate approaches, for example, density-functional tight binding, are much more inaccurate. Even though their lattice energies may be close to those obtained with density functionals\cite{jgb85},
their cell volumes and geometries are sometimes not even surpassing the accuracy of simple force-fields\cite{gad}.  
Many combinations of functionals and dispersion interactions have been used and developed with a special attention to
periodic systems\cite{mmort,mcut,erj,jnum,aot}, and in this contribution, we will evaluate some of them. In general, PBE\cite{pbe} with the D3 correction method of Grimme\cite{dftd3} and PBE with Many-body dispersion energy method\cite{mbd,mbdI} are viewed as some of the most
accurate functional and dispersion interaction combinations. Most of the time, computer codes utilizing plane waves such as
 VASP\cite{vasp}, CASTEP\cite{castep}, QUANTUM ESPRESSO\cite{quanex} or CP2K\cite{cp2k} are employed. Using codes with local basis
functions seems to be deprecated, as the use of diffuse functions will lead here to convergence problems.
For the the correct description of intermolecular interactions, however, the utilization of diffuse functions is extremely important.
For example, for hydrogen bonds, which are commonly found in molecular crystals, we were able to show that even employing of a
standard basis set of triple-zeta quality without diffuse functions will lead to an error of more than 2 kJ/mol on average for each dimer
calculated\cite{bosecppc} compared to the DFT basis set limit. For periodic codes with local (Gaussian) basis sets, even these "normal" basis sets of triple-zeta quality are
usually stripped
of their diffuse functions\cite{pient} for the code to converge. Counterpoise corrections will unfortunately not decrease these errors\cite{klopper}, at least when hydrogen bonds are concerned. Since we have many such bonds in typical molecular crystals, we expect the errors to
add up and yield inaccurate
results, unless there are error cancellation effects between the functional or dispersion correction and the incomplete basis set. Such effects, however,
are not systematic, as the best functional, even when developed for small basis sets, is the one computed at the basis set limit\cite{mhandy}.
This finally leaves us with density functionals which are restricted to the (meta)-generalized gradient approximation (GGA) type, as the
computation of hybrid functionals with
plane waves becomes easily prohibitively expensive. Despite their non-favorable scaling, hybrid functionals are used for molecular crystals as there is some indication that hybrid
functionals such as PBE0+MBD\cite{amr} may be more accurate than the commonly used PBE+MBD\cite{marom,schtuk}. We also experienced that especially when
computing different molecular conformers\cite{etylacet} for molecular crystals, hybrid functionals are more accurate than GGA ones. When computing a large amount of such conformers, this appears to be the general trend rather than an exception\cite{pantazis,manoj,karton,rezac,gruzman}.
To complicate things further, the application of post-Hartree-Fock methods for molecular crystals is still out of reach, despite recent progress\cite{masch,delbenm,booth,suhai}. If hybrid functionals using Hartree-Fock in combination with plane wave basis sets are computationally too demanding,
post-Hartree-Fock methods will be even more expensive. Furthermore, currently, no post-Hartree-Fock analytical gradients have been reported
for periodic systems.

An alternative approach is the use of embedding techniques. Most notably, MP2\cite{beran,beranmp2,schmidt} and other post-Hartree-Fock methods\cite{benan,fang,nolan,cervinka} have been embedded into point charges or force fields. Also closely related is the incremental method\cite{stoll,paulus,dollstoll,aherm,rosci,herm,jfried,jfriedr,hanrat,mull}, which uses a hierarchical scheme of 
Hartree-Fock and post-Hartree-Fock methods. Most of these
schemes have been employed to compute energies, but not geometries nor lattice parameters. Finally, Beran and co-workers introduced a method of
embedding either density functional theory, MP2, or post-Hartree-Fock methods into force fields, with the high-level
method computing all dimers given within a certain distance\cite{beran,beranqmmm,beranmp2}. Here, gradients and the gradients of lattice parameters are possible,
yielding an alternative to the above mentioned density functional theory plus dispersion corrections.\\
Very recently, we have published a series of papers of embedding MP2 into PBE+D2\cite{sauer} and BLYP+D3 into DFTB3+D3\cite{gad} using such dimer
interactions. This is our method of choice, as we can combine any molecular code with another one which utilizes periodic boundary conditions. Furthermore, we
have the advantage of a) using a very robust method as a low-level one for computing periodic structures b) getting away with a much smaller
distance for which the dimer contributions have to be calculated in comparison to force fields or point charges. Whereas the first point,
the more general applicability, may not be of as much concern as long as organic molecular crystals are computed, the second point makes 
our approach much faster. In general, we can reduce the number of dimers computed to a fraction when using much smaller cut-off distances making our
approach even competitive to the speed of DFT+D. Employing BLYP+D3:DFTB3+D3 in the embedding  scheme, we were able to show\cite{gad} that we basically 
 achieve the accuracy of the parent BLYP+D3\cite{becke,lyp} method.
The aim of this paper is two-fold:
1. We present the equations to compute an updated stress tensor which is used in a full optimization not only of the atomic coordinates, but also of the
lattice parameters.
2. We introduce two new embedding methods as alternatives to the commonly used DFT with hybrid functionals. We can compute B3LYP:BLYP and PBE0:PBE at a fraction of the
computational cost needed for hybrid functional calculations using plane wave codes, obtaining basically the same accuracy. There is also a direct link of PBE0:PBE to
the frequently employed HSE functional\cite{hse03,hse06} which screens the Hartree-Fock exchange of the hybrid PBE0 functional to zero for larger distances.\\
To compare all computational approaches, we utilize one of the most popular benchmark sets available for molecular crystals,
the X23 set of Reilly and Tkatchenko\cite{amr}. It is an extension
of the C21 set of Otera-de-la-Roza and Johnson\cite{aot}, which corrects even some lattice energies by using experimental heat capacities. It comprises of
several molecular crystals which are stabilized by van der Waals and/or hydrogen bonded interactions.
Of the X23 set, we took a subset of molecular crystals in order to optimize the lattice parameters of cubic, tetragonal and hexagonal cells
numerically using the BLYP+D3:DFTB+D3 embedding as reported earlier\cite{gad}, and compared these results to the analytically optimized lattice parameters obtained with the newly implemented stress tensors.
 We compare the obtained lattice energies and cell
volumes of PBE0 and B3LYP to those obtained with the embedded methods B3LYP:BLYP and PBE0:PBE.\\\\
{\bf{Performance of dispersion-corrected DFT and its approximative variants for the X23 set of molecular crystals}}\\
Table \ref{mean} summarizes the up-to-date reported mean absolute errors of lattice energies and cell volumes of X23 molecular crystals for different theoretical models.
\begin{widetext}
\begin{center}
\begin{table} [h!]
\caption{\label{mean} Mean absolute errors (MAE) of lattice energies $E$ (kJ/mol) and cell volumes $V$ ($\AA^3$) of X23 molecular crystals  with respect to experiment calculated with different theoretical models reported in the literature. }
\begin{tabular}{ccc}
\hline\hline
Method & MAE($E$) & MAE($V$)\\
\hline
\hline
DFTB3\cite{dftb3}+D3\cite{dftd3} & 13.0$^a$\cite{mmort},  13.3$^a$\cite{gad} & 36.6\cite{mmort}, 48.0$^a$\cite{gad}\\
DFTB3\cite{dftb3}+TS\cite{dftts} & 12.3$^a$\cite{mmort} & 17.8 \cite{mmort}\\
DFTB3\cite{dftb3}+MBD\cite{mbd,mbdI} & 20.3$^a$\cite{mmort} & 14.8 \cite{mmort}\\
HF-3c\cite{hf3c} & 8.2\cite{mcut} & 27.0 \cite{mcut} \\
PBE\cite{pbe}+D2\cite{dftd2} & 7.8\cite{jnum,aot} & \\
PBE\cite{pbe}+D3\cite{dftd3} & 4.0\cite{mmort}, 4.6 \cite{jmoel}, 5.0$^b$\cite{jmoel}, 5.5$^a$\cite{gad} & 6.4\cite{mmort}, 6.5 $^a$\cite{gad} \\
RPBE\cite{rpbe}+D3\cite{dftd3} &5.2$^a$\cite{gad} & 9.4$^a$\cite{gad}\\
PBE\cite{pbe}+LRD\cite{lrd} &  11.5\cite{yik} & 8.6$^c$\cite{yik}\\
revPBE\cite{revpbe}+LRD\cite{lrd} & 6.0\cite{yik} & 14.0$^c$\cite{yik}\\
PBE\cite{rpbe}+MBD\cite{mbd,mbdI} & 4.8\cite{mmort}, 5.7\cite{tgoul}, 5.9\cite{tbuc}, 6.2$^a$\cite{gad} &5.4$^a$\cite{tgoul}, 5.5\cite{mmort}, 5.6\cite{tbuc}, 5.9$^a$\cite{gad}\\
PBE\cite{pbe}+MBD\cite{mbd,mbdI}/FI$^d$\cite{tgoul} & 5.2\cite{tgoul}& 6.4$^a$\cite{tgoul}\\
PBE\cite{pbe}+TS\cite{dftts} & 13.4\cite{amr}, 14.0\cite{mmort}, 15.7$^a$\cite{gad} & 8.3\cite{mmort}, 9.1$^a$\cite{gad}\\
PBE\cite{pbe}+XDM\cite{bj,bjI,bjIII} & 4.6\cite{erj},6.5\cite{jnum,aot} & 8.2\cite{erjh}\\
BLYP\cite{becke,lyp}+XDM\cite{bj,bjI,bjIII} & 5.5\cite{erj}& 28.7\cite{erjh}\\
BLYP\cite{becke,lyp}+D3\cite{dftd3} & 9.9$^a$\cite{gad} & 17.0$^a$\cite{gad}\\
BLYP\cite{becke,lyp}+D3\cite{dftd3}:DFTB3\cite{dftb3}+D3\cite{dftd3} & 11.4$^a$\cite{gad} & 17.8$^a$\cite{gad}\\
PW86PBE\cite{pw86,pbe}+XDM\cite{bj,bjI,bjIII} & 3.7\cite{erj}& 7.8\cite{erjh}\\
B86bPBE\cite{b86b,pbe}+XDM\cite{bj,bjI,bjIII} & 3.6\cite{erj}&  8.5\cite{erjh}\\
optB88-vdW\cite{optB88} & 21.0$^a$\cite{gad} &14.2$^a$\cite{gad}\\
vdW-DF2\cite{df2} & 6.5$^a$\cite{gad} & 8.4$^a$\cite{gad}\\
TPSS\cite{tpss}+D3\cite{dftd3}/PAW\cite{paw} & 3.8\cite{jmoel}, 4.6\cite{sgr}, 5.0$^b$\cite{jmoel} & 8.5$^a$\cite{sgr}\\
B3LYP\cite{b3lypI,b3lypII}+D$^*$\cite{civa}/TZP\cite{tzp} & 4.6\cite{mcut}& 10.1\cite{mcut} \\
\hline\hline 
\end{tabular}
\begin{flushleft}
\hspace*{-0.52cm}$^a$Calculated based on data from supplementary material.\\
$^b$With the three-body contribution included. \\
$^c$For the C21 set.\\
$^d$FI denotes "fractional ions" (in a point-dipole dispersion approximation).\\
\end{flushleft}
\end{table} 
\end{center}
\end{widetext}

We deliberately excluded purely empirical (force field) methods and the methods with a high degree of empiricism (except for HF-3c) from Table \ref{mean}. Instead, we mostly focus on the data obtained with dispersion-corrected (+D) DFT and DFTB. All data in Table  \ref{mean} correspond to the fully lattice-optimized X23 structures calculated with a given model whereas Table S1 (see supplementary material) also reports single-point results based on either experimental or calculated at a lower-level lattice parameters.\\
One can easily see that highly-approximative methods (DFTB+D and HF-3c) lead to significant errors in both lattice energies and cell volumes (with the respective minimal mean absolute errors (MAEs) of at least 8.2 kJ/mol and 14.8 $\AA^3$ for lattice energy and cell volume, respectively). Among DFT+D approaches, the PBE functional combined with different dispersion models (empirical D3, many-body dispersion MBD) provides a rather accurate description of both lattice energies and cell volumes of the X23 crystals with typical MAEs of 4-5 kJ/mol in energies and 5.5-6.5 $\AA^3$ in cell volumes. We note that not all dispersion models are equally applicable to PBE as local response dispersion (LRD), Tkatchenko-Sheffler (TS), exchange-hole dipole moment (XDM) models lead to much larger errors (indicating the lack of proper parameter adjustment for the specific functional and/or methodological issues). On the other hand, the BLYP functional with D3 model leads to twice larger errors compared to PBE+D3 ones: $\sim$ 10 kJ/mol in lattice energies and 17 $\AA^3$ in cell volumes. The former error can be reduced to 5.5 kJ/mol when employing XDM for dispersion, however, the respective error in cell volume increases then to 28.7 $\AA^3$. This is a clear indication that one cannot achieve the best performance of both lattice energies and cell volumes with the same set of dispersion parameters within a given DFT+D approach. As has been mentioned earlier, the BLYP+D3 accuracy can be achieved utilizing the fragment-based BLYP+D3:DFTB+D3 method with a suitable choice of the trust radius (or cutoff) parameter responsible for the dimer selection\cite{gad}.
Among the most accurate DFT+D candidates, PW86PBE+XDM, B86bPBE+XDM and  TPSS+D3/PAW lead to the smallest errors: 3.6-3.8 kJ/mol in lattice energies and 7.8-8.5 \AA$^3$ in cell volumes (cf. Table 1). On the other hand, dispersion-corrected PBE errors typically lie within 4.0-6.5 kJ/mol for the lattice energies and 5.4-8.2 \AA$^3$ for the cell volumes (except for D2, LRD and TS dispersion models), i.e. they provide even slightly more accurate optimized cell volumes than the ones stemming from the former set of hybrid and meta-GGA functionals.

That is why PBE+D3 and PBE+MBD are probably the most utilized functional/dispersion combinations used in this field.


\section{\sffamily \Large Computational Methods}
\subsection{Periodic and Molecular Calculations}
The periodic calculations of molecular crystals were performed with VASP program\cite{vasp}, employing a large 1000 eV cutoff and hard potentials. 
All QM:QM calculations have been performed on the standard k-points grid except calculations involving MBD dispersion for which the large set of k-points have been applied (see Table S2).\\
As DFT functionals we used: PBE\cite{pbe} and BLYP\cite{becke,lyp} and the corresponding QM:QM methods to mimic hybrid PBE0\cite{pbe0} and B3LYP\cite{b3lypI,b3lypII}. 
 We applied the following dispersion corrections: the many-body dispersion energy method (MBD) of Tkatchenko et al. \cite{mbd,mbdI} and the DFT+D3 dispersion correction methods of Grimme et al.\cite{dftd3} with Becke-Johnson damping method\cite{bj,bj0,bjI,bjII}. 
Calculations of molecular fragments at high and low levels were performed with TURBOMOLE (TM) program\cite{turbo}, using def2-TZVPPD\cite{def2} basis set. 
For the comparison between numerically and analytically optimized lattice and energetic parameters the BLYP+D3:DFTB3+D3 model\cite{gad} has been employed. In this model the low-level calculations have been performed using density functional tight binding in its third-order expansion of the Kohn-Sham energy with respect to charge density fluctuations\cite{dftb3} along with the 3ob set of Slater-Koster parameters\cite{dftbIII} as it is implemented in the DFTB+ program\cite{dftbI,dftbII}. 
The optimisations of the molecular crystal structures was carried out by modified version of the QMPOT program\cite{qmpot}. The total energy was calculated according to the following form:

\begin{equation}
\hspace*{-2cm}
\begin{split}
 E_{QM:QM}=E_{Low-level}(host)-\sum_i^{N=Dimers}[E_{Low-level}^{Cluster}(i)-E_{High-level}^{Cluster}(i)]-\\
\sum_j^{N=Monomers\in Dimers}[E_{Low-level}^{Cluster}(j)-E_{High-level}^{Cluster}(j)]+\\ 
\sum_k^{N=Monomers\notin Dimers}[E_{Low-level}^{Cluster}(k)-E_{High-level}^{Cluster}(k)]
\label{eq:qmp}
\end{split}
\end{equation}
The first term of Eq. (\ref{eq:qmp}), $E_{Low-level}(host)$, implies periodic computations of molecular crystal in the space of arbitrary chosen $k$-points. The $E_{Low-level}^{Cluster}$ and $E_{High-level}^{Cluster}$ represent calculations of dimers or/and monomer energies at low and high levels, respectively. The same scheme has been adapted for gradients. The number of dimers can be tuned by the trust radius - a threshold distance around the constituent monomer fragments. In our calculations we set the trust radius parameter value to 4 \AA.
To get a fully relaxed structure, the optimisation of atom positions and cell lattice parameters is
required. In this work, we report the implementation of a cell optimisation within this QM:QM approach.
 Here, we combine the analytical stress tensor from periodic calculations with the corresponding stress
originating from the cluster contributions involving one periodic image fragment. The calculation of the stress tensor contrasts a previous contribution of Beran and co-workers, who introduced the calculation of cell gradients to such embedding methods\cite{beranmp2}. The advantage of a stress tensor vs a cell gradient is that intramolecular forces are taken into account for the former, making the overall optimization faster.

\subsection{Stress Tensor}
The variation of atomic positions including lattice vector endpoints upon lattice change can be described according to the following relationship\cite{angy}:
\begin{align}
v_{i}^{\prime}=\sum_{j}(\delta_{ij}+\epsilon_{ij})v_{j}
\end{align}

where $\delta_{ij}$ is the Kronecker delta and $\epsilon_{ij}$ is the element of the the symmetrical strain tensor\cite{doll}:
  
\begin{align}
\begin{pmatrix}  \epsilon_{11} & \epsilon_{12}& \epsilon_{13}\\ 
                 \epsilon_{12} & \epsilon_{22}&\epsilon_{23} \\
                 \epsilon_{13} & \epsilon_{23}&\epsilon_{33} \\
\end{pmatrix}
\end{align}

 
The stress tensor used to optimize the unit cell of crystal structure can be computed from the total energy as\cite{doll,knuth,angy}:

\begin{align}
\sigma_{i,j}=-\frac{1}{V}\frac{\partial E}{\partial \epsilon_{ij}}
\end{align}

where $V$ is the volume of unit cell:
\begin{align}
V=abc\sqrt{1 - cos^2\alpha - cos^2\beta - \cos^2\gamma + 2cos\alpha cos\beta cos\gamma}
\end{align}
  $a$, $b$, $c$, $\alpha$, $\beta$, $\gamma$ are lattice constants and angles. 
The stress tensor depends only on the lattice vectors ($v$) and on the atomic positions ($R$). Hence, in addition to explicit lattice vector dependence of the total energy, atomic forces contribute also to the strain-derivative tensor. For a given strain component $\epsilon_{ij}$, the lattice vector derivatives of the energy from the non-periodic fragment calculations can be evaluated as\cite{knuth}:

\begin{align}
\frac{\partial E}{\partial \epsilon_{ij}}=\sum_{k}\frac{\partial E}{\partial R_{k,i}}R_{k,j}+\sum_{n}\frac{\partial E}{\partial v_{in}}v_{jn}
\label{eq:stress}
\end{align}

where $v$ defined in terms of three-dimensional vectors {\bf{v$_1$}},  {\bf{v$_2$}},  {\bf{v$_3$}} in a six-parameter convention form:

\begin{align}
\begin{pmatrix} v_1\\
v_2\\
v_3\\
\end{pmatrix}=\begin{pmatrix} a & 0 &0\\
bcos\gamma & bsin\gamma& 0\\
ccos\beta & c\frac{cos\alpha-cos\beta cos\gamma}{sin\gamma} &\frac{V}{absin\gamma} \\
\end{pmatrix}
\end{align}

 The term  $\frac{\partial E}{\partial R}$ in eq.\ref{eq:stress} represents atomic forces (High-Low):

\begin{align}
\frac{\partial E}{\partial R}=\frac{\partial E^{Cluster}_{Hi-level}}{\partial R}-\frac{\partial E^{Cluster}_{Low-level}}{\partial R}
\end{align}

 and each  $\frac{\partial E}{\partial v}$ term is given by\cite{beranmp2}:
\begin{widetext}
\begin{equation}
\frac{\partial E^{High,Low}}{\partial v}=
\sum_{i}\sum_{j}\sum_{k}
 \bigg\{ \bigg(\frac{\partial E_{ij}^{High}}{\partial R_{k}}-
\frac{\partial E_{j}^{High}}{\partial R_{k}} \bigg)-
 \bigg(\frac{\partial E_{ij}^{Low}}{\partial R_{k}}-
\frac{\partial E_{j}^{Low}}{\partial R_{k}} \bigg) \bigg\}\;. \label{eq:wideeqn}
\end{equation}
\end{widetext}

The index $i$ sums over monomers in the central unit cell, $j$ sums over periodic image monomers which lie within the trust radius distance of $i$th monomer, and $k$ sums over the $k$th atom in the image monomer $j$. It should be noted that the use of Cartesian coordinates instead of fractional coordinates  means that monomers and dimers that lie entirely within the central unit cell do  not contribute to the lattice parameter gradient terms.  Therefore, only dimer terms involving one periodic image molecule have a non-zero contribution.

Thus, the total lattice-vector derivatives of the energy are the sum of the analytical derivatives from periodic calculations and the gradients of the energy obtained from the cluster contributions:
\begin{align}\label{eq:etot}
\frac{\partial E^{QM:QM}}{\partial  \epsilon_{ij}}=\frac{\partial E^{QM}_{PBC}}{\partial  \epsilon_{ij}}+ \frac{\partial E^{QM}}{\partial  \epsilon_{ij}}
\end{align}

\subsection{Optimization of the Unit Cell Parameters}
The optimization of cell parameters used in this study is based on a conjugate gradient algorithm\cite{numrecipes}.
At the initial optimization cycle, the trial steepest descent step (i.e. in the direction of the stress tensor $\sigma$) is performed: 
\begin{align}
\omega=\frac{\sigma_k+\sigma_{k-1}\sigma_k}{\sigma_{k-1}\sigma_{k-1}}
\end{align}
Then, the energy and the cell gradients are recalculated according to the conjugate gradient algorithm which requires the line minimization along search directions ($s_{k+1}$):
\begin{align}
s_{k+1}=\sigma_k+\omega s_k
\end{align}
After that step the stress and energy are recalculated again. If the stress tensor contains a significant component parallel to the previous search direction, then the line minimization is improved by further corrector steps using a variant of Brent's algorithm\cite{numrecipes}.\\
We adopt the scheme in which the geometry is optimized on the basis of respective energies and gradients (as described in Section III A), at fixed $a$, $b$, $c$, $\alpha$, $\beta$, $\gamma$ parameters during microiterations whereas the cell is relaxed during macroiterations. When the convergence criteria on energy and gradients is achieved during the last step of microiterations, the analytical cell gradients from low-level (host) calculations are combined with the corresponding contributions of cluster calculations (see eq. \ref{eq:etot}). Then, the total stress of the cell is  given to the optimizer to generate the new cell vectors. Within the conjugate gradient optimizer we use fractional coordinates in conjunction with direct lattice vectors. The conversion of Cartesian coordinates into fractional is performed in the following way:
\begin{widetext}
\begin{align}
\begin{pmatrix} h_1\\
..\\
h_N\\
\end{pmatrix}
=\begin{pmatrix} \frac{1}{a} & 0 &0\\
-\frac{cos\gamma}{asin(\gamma)} & \frac{1}{bsin\gamma}& 0\\
bc\frac{cos(\alpha)cos(\gamma)-cos(\beta)}{Vsin(\gamma)} & ac\frac{cos(\beta)cos(\gamma)-cos(\alpha)}{Vsin(\gamma)} &ab\frac{sin(\gamma)}{V} 
\end{pmatrix}\begin{pmatrix} x_1 & y_1 & z_1\\
.. & .. & ..\\
 x_N & y_N & z_N\\
\end{pmatrix}
\end{align}
\end{widetext}
where $h_N$ are the fractional coordinates of the $N\text{-}th$ atom.

\subsection{Dissociation energy calculations}
In our first contribution on this subject\cite{sauer}, $E^{cluster}_{low-level}$ of eq. \ref{eq:qmp} was calculated by utilizing periodic VASP in a large unit cell. Because we have to compute a large number of dimer fragments and atoms, this would become the time-determining step for our QM:QM calculations with hybrid and GGA functionals. When employing BLYP+D3:DFTB3+D3\cite{gad}, this is not an issue, since DFTB3+D3 can be computed for both periodic as well as molecular structures on the same footing, i.e. with the same code. For our target PBE0:PBE+D3, however, we would have to perform every PBE+D3 dimer fragment step in a large unit cell, leading to convergence issues and a slow performance. Thus, the PBE+D3 values in the gas phase are computed by the molecular code using a sufficiently large def2-TZVPPD basis set. For the gradients and thus geometries the two approaches essentially yield the same results, since we can show these to be equal for both PBE+D3/TZVPPD and PBE+D3/PAW (see Table S3). However, we have to change the calculation of the dissociation energies, since in this case the low-level calculations are not performed on the same footing. We do this by correcting only the monomer contributions.  
The dissociation energy ($DE$) of molecular crystal is  thus calculated as:
 \begin{align}
DE=(NE^{Monomer}-E_{QM:QM})/N
\end{align}
where $N$ is the number of monomers and $E^{Monomer}$ is the energy of monomer calculated as the following:
 \begin{align}
E^{Monomer}=E^{Monomer}_{High-level}-E^{Monomer}_{Low-level}+E^{Monomer}_{Low-level}(Host)
\end{align}
where the last term is calculated at the  gamma point in VASP with the large cell volume (20x20x20 \AA) and small van der Waals radius (10 \AA).
The high-level optimized geometry is used in all low-level monomer energy calculations. 
This way, we speed up a computing time of the molecular crystal by a large amount without any loss in accuracy.
\section{\sffamily \Large Results and Discussion}
\subsection{Comparison of Analytical and Numerical Cell Volumes}
As mentioned in the introduction, we computed the variation of the sublimation energy with the lattice constant for all cubic, orthorombic, hexagonal and tetragonal molecular crystals of the X23 set both numerically as well as analytically. In numerical calculations, only  the atoms were relaxed in the frame of constant lattice parameters and those were varied manually. The results  are shown in Table \ref{annum}: we found that our optimized structures are in a very good agreement with equilibrium volume structures obtained by minimizing the series of constrained-volume relaxed crystals. Hence, our optimization algorithm validated by the numerical optimization procedure. The small difference between numerical and analytical values can be attributed to the absence of symmetry constraints on lattice parameters ($a$ $\neq$ $b$ $\neq$ $c$) in case of the analytical optimization. Another potential factor which may affect the lattice parameters optimization is the plane wave basis-set incompletness error\cite{gpf}. However when the high cutoff energy of 1000 eV is applied, the so-called "Pulay stress" connected with this problem is considered to be negligible.
\begin{widetext}
\begin{center}
\begin{table} [h!]
\caption{\label{annum} Comparison between numerically and analytically optimized lattice and energetic parameters of cubic, tetragonal, hexagonal and orthorhombic X23 crystal phases obtained with BLYP+D3:DFTB3+D3 (trust radius = 5 \AA, basis set at high level: def2-TZVPPD).}
\scalebox{1.0}{\begin{tabular}{lcccc}
\hline\hline
           {\multirow{2}{*}{ X23 structure}}   & \multicolumn{2}{c}{Numerical optimization}& \multicolumn{2}{c}{Analytical optimization}    \\
                            & V, \AA$^3$ & E, kJ/mol &V, \AA$^3$ & E, kJ/mol \\
\hline\\
Acetic acid     &  290.8   &  72.76       & 291.36      & 72.75    \\
Adamantane& 370.4  & 79.45   & 366.70  & 79.29 \\  
Benzene&  441.64 &62.03   & 436.69  & 61.93 \\  
Carbon dioxide& 176.74  & 25.14  & 175.79  & 25.13 \\  
Cyanamide& 393.31  & 99.2  & 390.05  & 99.33 \\  
Cytosine& 448.21  & 177.24  & 448.73  & 177.23 \\  
Hexamethylentetramine& 322.24  &  95.61 &324.87   & 95.55 \\  
Ammonia& 114.08  & 44.47  & 114.88  & 44.42 \\  
Oxalic acid ($\alpha$) & 312.41  & 110.77  & 311.01  & 110.72 \\  
Pyrazine & 188.16  & 75.34  &  187.04 &  75.32\\  
Pyrazole & 676.14  & 84.34  &  675.48 & 84.33 \\  
$s$-Triazine &  536.48 & 68.93  &  532.8 & 68.88 \\  
$s$-Trioxane & 594.25  & 66.44  &  594.02 &66.44  \\   
Urea & 145.45  & 111.04  &  144.63 & 110.9  \\
\hline\hline\\
\end{tabular}} 
\end{table}
\end{center}
\end{widetext}
\subsection{Comparison of B3LYP:BLYP(B3LYP+D3), PBE0:PBE(PBE0+D3),  PBE0:PBE(PBE0+MBD) with periodic DFT}
With all these prerequisites of subsection IIA, we can compute the properties of molecular crystals with hybrid density functionals. It is important to note that we do not make any reparametrization  attempts of dispersion coefficients in this publication, like it would be the case for the somewhat related HSE functional. We use in the resulting PBE0:PBE+D3, PBE0:PBE+MBD, and B3LYP:BLYP+D3 the respective D3 and MBD dispersion coefficients from the parent hybrid method. However, this introduces some inconsistencies for the long-range part of the dispersion model, where the low-level GGA functional is now utilized together with the dispersion model of the hybrid functional. For the X23 set and BLYP:DFTB3+D3, the differences in lattice energies and cell volumes in comparison to BLYP+D3 were small, since the long range tails of BLYP and DFTB3 turned out to be remarkably similar for organic molecular crystals. \\
In addition, to introduce a method which should mimic hybrid functionals as close as possible, we actually need to compare to hybrid functional values. For this purpose, we additionally performed the computationally extremely expensive hybrid functional computations for the X23 set of molecules. To be able to  perform these calculations, we had to reduce the k-point sampling. Especially for the MBD method, using a high k-point grid may become crucial since the computation of dispersion is performed in reciprocal space\cite{tbuc}. 

\begin{figure}[!]
\begin{center}
\caption{\label{pbed3} Histograms showing an error distribution of (a) {\it{X23}} lattice energies and (b) {\it{X23-2}} cell volumes (see text) with respect to experiment using DFT+D3 and PBE0:PBE(PBE0+D3)}
\vspace{1.5cm}
\scalebox{0.75}{\includegraphics[angle=0,origin=c]{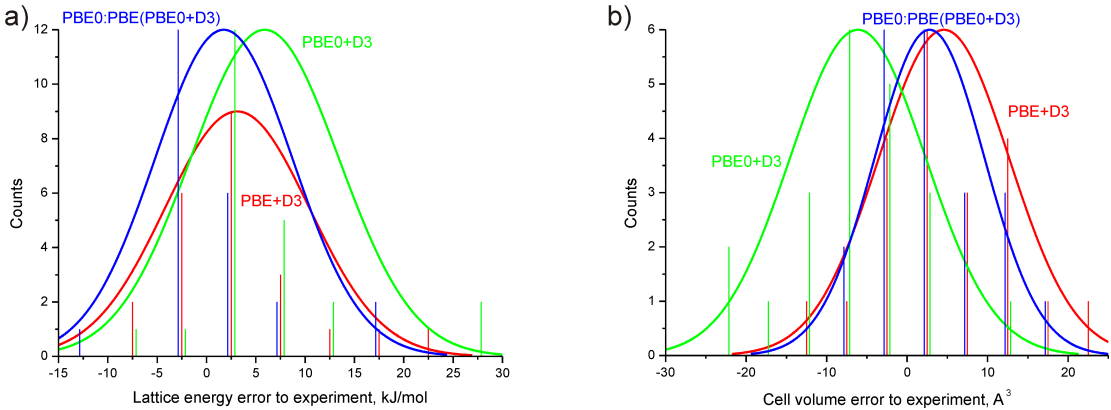}}
\end{center}
\end{figure}

\begin{figure}[!]
\begin{center}
\caption{\label{pbembd} Histograms showing an error distribution of (a) {\it{X23}} lattice energies and (b) {\it{X23-2}} cell volumes with respect to experiment using DFT+MBD and PBE0:PBE(PBE0+MBD)}
\vspace{1.5cm}
\scalebox{0.75}{\includegraphics[angle=0,origin=c]{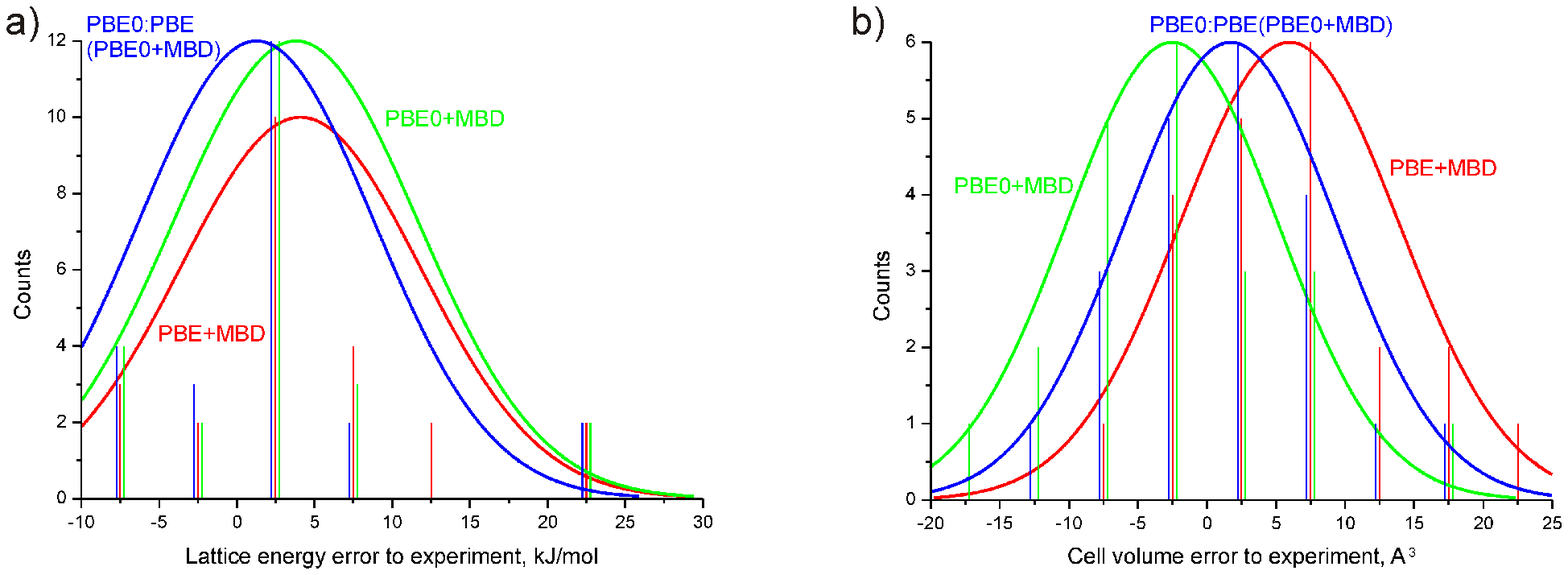}}
\end{center}
\end{figure}

\begin{figure}[!]
\begin{center}
\caption{\label{blypd3} Histograms showing an error distribution of (a) {\it{X23}} lattice energies and (b) {\it{X23-2}} cell volumes with respect to experiment using DFT+D3 and B3LYP:BLYP(B3LYP+D3)}
\vspace{1.5cm}
\scalebox{0.75}{\includegraphics[angle=0,origin=c]{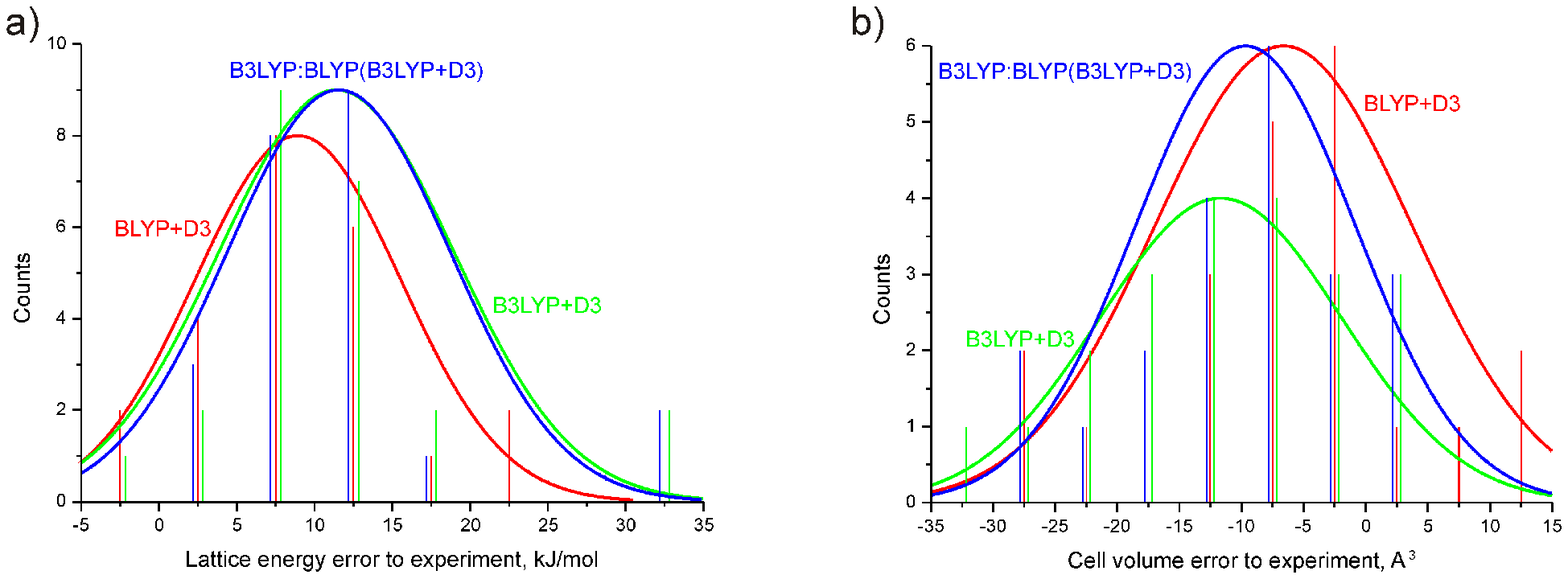}}
\end{center}
\end{figure}

Because of this, the PBE0+MBD functional is likely to have the largest dependence on the k-points of the  three hybrid functionals tested. In Table \ref{tab:table3},
we investigated the change in energies by  going to much larger k-point grids (see supplementary material Table S2,S4) and performed single-point calculations at the structures obtained from PBE0+MBD with the reduced  (small) grid size.  All lattice  energies with the larger grid are smaller, on average by 1.7 kJ/mol and by a maximum of 4.5 kJ/mol.
Compared to experiment, the lattice energies of the smaller grid yield an accuracy of 2\% and a maximum deviation of 3.5\%, reducing the mean absolute error of PBE0+MBD from 5.9 kJ/mol to 5.4 kJ/mol. 
For the cell volumes, only a small number of molecular crystals could be computed, indicating that the volumes are even accurate to 0.3\% for the small grid (see supplementary material Table S4).
 Of course, for our embedding method PBE0:PBE(PBE0+MBD), such a k-point variation is computationally rather inexpensive. \\
Having obtained the hybrid functional values, we can conclude from Table \ref{embed} and the histograms in Figures \ref{pbed3},\ref{pbembd},\ref{blypd3} that in fact, the hybrid functionals actually perform worse than their GGA counterparts regarding their RMS errors. This is somewhat surprising, since a large part of CPU time was spent previously on the computation of molecular crystals with hybrid functionals  such as PBE0+MBD\cite{amreilly,marom,schtuk}. Possibly, the better description of intramolecular interactions by hybrid functionals\cite{etylacet,pantazis,manoj,karton,rezac,gruzman} leads to a better prediction when the relative energies of several polymorphs are concerned.
In addition, Figures 1-3 reveal that the maximum of the distribution curves of the lattice energies are higher for the hybrid functionals than for their GGA counterparts, indicating that the errors of the hybrid functionals are more systematic.
This may be, besides the better description of the intramolecular interactions and conformers, another reason why hybrid functionals may be superior, at least when the energies of different polymorphs are concerned. This issue, however, has to be investigated further and more systematically possibly by means of the here introduced embedding method.
Also note that the energy differences between the GGA and hybrid functionals are not large. The errors could be shifted if thermal effects were considered for all molecular crystals of the X23 set. Here, we took the reference values of Mortazavi et al\cite{mmort}. In this paper, the reference energy values were back-corrected by vibrational corrections computed with PBE+TS, while in some cases, thermal corrections have been employed (see original X23 paper)\cite{amr}.\\
Concerning the cell volumes, thermal and especially zero-point effects, computed by the quasi-harmonic approximation, can have a large impact. To back-correct these values in a similar manner as it was done for the energies, we use the available data from the literature. Here, we utilize the force field data from Day and co-workers\cite{jnum} together with the CO$_2$ data of Beran and co-workers\cite{beranco} and the urea data of Civalleri and co-workers\cite{civall}. For hexamethylentetramine, a crystal structure with a much lower temperature 
 (34 K) has been utilized\cite{becka}, whereas the force field yielded a reduction of only 1.8\% at much higher temperatures\cite{jnum}. Hence, we considered the hexamethylentetramine structure at 34 K as a reference point. In general, the force fields agree well with some other computed data: for imidazole and acetic acid, the reduction is 2.7 \% and 1.6\% of the force field compared to 3.2\% and 2.2\% with MP2/TZ\cite{beranheit}, and the value for NH$_3$ is 5.8\% compared to 5.2\% for ND$_3$ calculated with PBE+MBD\cite{hoja}. For benzene the 2.4\% reduction predicted by the force field is close to the 2.2\% which was experimentally determined by the 4K neutron diffraction measurement\cite{rubl}. The new reference values are displayed in the supplementary material in Table S5, whereas the FIT force field of Day and co-workers\cite{jnum} appears to yield slightly better values than the ones utilizing the W99rev6311P5 force field, especially for NH$_3$ and benzene. Unfortunately, we had to exclude two systems with the largest thermal corrections, as these values appear unreasonable: pyrazole and s-triazine. Hence, we arrive at a $X23-2$ reference set which finally includes thermal corrections. By doing so, the errors of all functionals are reduced by a large amount, for PBE+MBD up to a factor of almost two (from 19.1 to 10.0 $\AA^3$ for the RMS error), indicating that these two values may have been outliers. The general conclusions are not affected by the ommittance of these two outliers (see Table S6 and discussion below). \\
In case of Table \ref{mean}, we can now recalculate some of the mean absolute errors of the cell volumes for the functionals for which we have all data points (see Table S7 of the supplementary material). The errors of PBE+TS, BLYP+D3 and optB88-vdW are reduced, while the errors for the other GGA functionals reported were increased when going from the non-thermally corrected to the thermally corrected reference values. The same holds for the GGA functionals in Table \ref{embed}. Whereas the cell volume was underestimated slightly by PBE+D3 and PBE+MBD when comparing it to the uncorrected cell volumes, it is overestimated by the corrected ones. And while PBE+D3 and PBE+MBD appear more accurate without thermal and quasi-harmonic corrections, PBE0+D3 and PBE0+MBD are more accurate with them. BLYP+D3 and B3LYP+D3 yield, including corrections, much lower errors which are now comparable to the PBE values when the \% errors are considered. 
 If we would look at the whole X23 set of molecules and their thermal corrections, the same conclusions hold. The RMS errors change from 2.9\%/9.3$\AA^3$ (X23) to 2.8\%/7.0$\AA^3$ (X23-2) for PBE+D3 without thermal correction and from 4.6\%/16.3$\AA^3$ (X23) to 4.1\%/9.3$\AA^3$ (X23-2) with thermal correction. 
The PBE+MBD values change from 2.9\%/8.1$\AA^3$ (X23) to 2.9\%/7.1$\AA^3$ (X23-2) without and from 5.1\%/19.1$\AA^3$ (X23) to 4.5\%/10.0$\AA^3$ (X23-2) with thermal correction.
For PBE0+D3, the corresponding results are 5.0\%/18.7$\AA^3$ (X23) to 4.7\%/15.5$\AA^3$(X23-2) without and 3.5\%/12.2$\AA^3$(X23) to 3.3\%/9.9$\AA^3$ (X23-2) with thermal correction of the experimental reference.
For all cell volumes, the errors of the QM:QM scheme methods are ubiquitously in between the GGA functionals and their respective hybrid functional counterparts.\\
This can be seen by the values of Table \ref{tab:dftqm}, where the \% RMS deviations between PBE0+D3 and PBE+D3 to our PBE0:PBE+D3  method are shown.  Here, the PBE0:PBE+D3 values are close to PBE+D3, as indicated by the results displayed.
In contrast, the PBE0:PBE+MBD values are much closer to PBE0+MBD. For B3LYP:BLYP+D3, a similar picture emerges, as the RMS errors to BLYP+D3 and B3LYP+D3 are 3.8 and 1.7 kJ/mol,  respectively. One reason for this obvious difference is the flat potential energy surface of the PBE+D3 and PBE+MBD methods as well as their hybrid counterparts. As can be seen from Table S10 in the supplementary material, where we performed single-point calculations of our PBE0:PBE+D3 method on both the PBE0+D3 and PBE+D3 structures, the energy differences for these, similar  geometries are very small. Thus, the total sum of RMSD between QM:QM and PBE0+D3, PBE+D3 geometries obtained from Kabsch algorithm\cite{kabsch,kabsch2} within ArbAlign code\cite{arbalign} are 0.85 and 0.54$\AA$ correspondingly.
The slightly higher lattice energy especially for the PBE0:PBE+MBD values compared to that of the PBE0+MBD structures are due to the k-point grid used for the embedded calculations.  Especially the single-point values at the PBE0 and B3LYP structures are extremely close to the optimized ones (0.2 kJ/mol on average), indicating that the potential energy surface of PBE0:PBE is basically flat in that region. It explains the difficulty of PBE0:PBE+D3 to capture the cell volume of the hybrid methods  (here, the deviation is much larger than for lattice energies). This implies that for B3LYP+D3, where the dispersion part is much larger, the energy differences are less determined by the functional part and the potential energy surface is steeper. 
For PBE0+D3 or PBE+D3, the short-range dispersion of the functional appears to play a larger role and thus, the embedded value is closer to PBE+D3.  The MBD dispersion in PBE is slightly larger than D3, therefore the PBE0:PBE+MBD values are closer to PBE0+MBD, in contrast to D3.\\

Thus, B3LYP:BLYP+D3 is very close to B3LYP+D3, whereas PBE0:PBE+MBD are in between PBE and PBE0. As before, PBE0:PBE appears somewhat closer to PBE+D3 and PBE0:PBE+MBD to PBE0+MBD. \\ 
Although we are able to reproduce B3LYP+D3 rather well with B3LYP:BLYP+D3, the PBE0:PBE+D3 and PBE0:PBE+MBD methods may need some reparametrization of the dispersion coefficients, as it was done for the HSE+D3 method.
Nevertheless, for the monomers and dimers, we automatically achieve the accuracy of hybrid functionals for the introduced method, which may be an important step when comparing different polymorphs.  
\begin{center}
\begin{table} [h!]
\caption{\label{tab:table3}PBE0+MBD: X23 lattice energies calculated with higher and lower k-point grids and respective errors to experiment.}
\begin{tabular}{lccccc}
\hline
\hline
 X23 phase           &     Small & Large &  Exp &  Small-Exp &  Large-Exp\\
\hline
 Acetic acid           &    75.64     &     74.20   &  72.8   &   2.84     &       1.40       \\
 Adamantane            &    75.49     &     73.90   &  69.4   &   6.09     &       4.50       \\
 Anthracene            &   106.65     &    105.13*  &  112.7  &   -6.05    &       -7.57      \\
 Benzene               &    54.52     &     52.82   &  55.3   &  -0.78     &      -2.48       \\
 Carbon dioxide        &    23.02     &     22.80   &  28.4   &  -5.38     &      -5.6        \\
 Cyanamide             &    88.90     &     88.08   &  79.7   &    9.2     &       8.38       \\
 Cyclohexane-1,4-dione &    93.42     &     90.29*  &   88.6  &    4.82    &        1.69      \\
 Cytosine              &   159.42     &    157.80*  &  162.8  &   -3.38    &       -5.00      \\
 Ethyl carbamate       &    90.09     &     87.30   &  86.3   &   3.79     &       1.00       \\
 Formamide             &    81.21     &     80.17   &  79.2   &   2.01     &       0.97       \\
 Hexamethylenetetramine&    86.49     &     85.34   &  86.2   &   0.29     &      -0.86       \\
 Imidazole             &    90.50     &     89.10   &  86.8   &    3.7     &       2.30       \\
 Naphthalene           &    79.56     &     76.82   &  83.1   &  -3.54     &      -6.28       \\
 Ammonia               &    40.45     &     40.27   &  37.2   &   3.25     &       3.07       \\
 Oxalic acid ($\alpha$)&   120.25     &    118.04   &  96.3   &  23.95     &      21.74       \\
 Oxalic acid ($\beta$) &   120.49     &    118.58   &  96.1   &  24.39     &      22.48       \\
 Pyrazine              &    60.13     &     58.77   &  61.3   &  -1.17     &      -2.53       \\
 Pyrazole              &    80.24     &     78.20*  &   77.7  &    2.54    &        0.50      \\
 Succinic acid         &   135.05     &    133.16   & 130.3   &   4.75     &       2.86       \\
 s-Triazine            &    55.78     &     54.02*  &   61.7  &   -5.92    &       -7.68      \\
 s-Trioxane            &    60.93     &     59.30*  &   66.4  &   -5.47    &       -1.63       \\
 Uracil                &   139.54     &    135.00*  &  135.7  &    3.84    &       -0.70      \\
 Urea                  &   111.84     &    111.55   & 102.5   &   9.34     &       9.05       \\
\hline
\hline
\end{tabular}
\begin{flushleft}
* Calculated with a standard (GGA) set of k-points (for its definition - see Table S1).
MAE(Small) = 5.93 kJ/mol\\
MAE(Large) = 5.29 kJ/mol
\end{flushleft}
\end{table}
\end{center}

\begin{widetext}
\begin{table} [h!]
\begin{center}
\caption{\label{embed} Lattice energy (kJ/mol) and cell volume ($\AA^3$)  errors of complete X23 and truncated (X23-2)$^{*}$ crystal sets computed by embedded QM:QM and  pure QM methods with respect to experimental reference values with Ref$_{th}$ and without (Ref.) thermal expansion correction$^{**}$ applied.}
\end{center}
\begin{center}
\hspace*{-1.7cm}
\scalebox{0.75}{\begin{tabular}{lcccccc}
\hline\hline
 method &lattice energy error &lattice energy \% error &cell volume error &cell volume \% error       &cell volume error &cell volume \% error\\
 (dispersion)       & RMS /\ MAE   /\ mean  & RMS  /\ mean  & RMS /\  MAE   /\ mean & RMS /\ mean     & RMS /\  MAE   /\ mean & RMS /\ mean\\
                    &       \multicolumn{2}{c}{Ref. X23}                                              &   \multicolumn{2}{c}{Ref. X23-2}           & \multicolumn{2}{c}{Ref$_{th}$ X23-2} \\      
\hline
 PBE+D3             & 7.8 /\ 5.3 /\   3.1   &9.6  /\ 3.4    &7.0  /\  5.2  /\ -3.6  &2.8 /\ -1.1      &9.3 /\  7.0  /\ 4.7  &4.1 /\  1.9\\
 PBE0:PBE(PBE0+D3)  & 7.0 /\ 4.1 /\   1.8   &8.3  /\ 1.3    &8.1 /\  6.4  /\ -5.3   &3.0 /\ -1.7      &7.5  /\ 5.6  /\ 2.9  &3.7 /\  1.3\\
 PBE0+D3$^{***}$        & 9.4 /\ 6.8 /\   5.9   &10.7 /\ 6.4   &16.2 /\  14.3 /\ -14.3 &4.9 /\ -4.4   &10.4 /\  8.4 /\ -6.0 &3.4 /\ -1.5\\
\hline
 PBE+MBD\cite{gad}  & 8.5 /\ 6.3 /\   3.5   &10.5 /\ 3.6    & 7.1 /\   4.6  /\  -2.2 &  2.9 /\ -0.7   &10.0 /\   7.7  /\  6.1&  4.5 /\ 2.4\\
 PBE0:PBE(PBE0+MBD) & 7.6 /\ 5.1 /\   0.6   & 9.5 /\-0.3    &9.8 /\   7.3  /\  -6.4 &  3.3 /\ -2.0    &7.9 /\   6.3  /\  1.8 &  3.5 /\ 1.0  \\
 PBE0+MBD$^{***}$       & 8.8 /\ 5.9 /\   3.2   &10.1 /\ 2.9   &13.1 /\   11.1 /\ -10.6 & 4.1 /\ -3.3 &8.1 /\  6.5 /\ -2.4  &  3.4/\ -0.4  \\ 
\hline
 BLYP+D3            &11.0 /\  9.3/\   8.9   &13.0 /\ 10.3   &17.6 /\  14.8 /\ -14.8 & 5.0 /\ -4.5     &12.2 /\  9.7 /\  -6.5& 3.6 /\ -1.5 \\
 B3LYP:BLYP(B3LYP+D3) & 13.6 /\ 11.6 /\ 11.6&15.6 /\ 13.7   &19.9 /\  17.8 /\  -17.8& 5.8 /\ -5.6     &13.0 /\  10.8 /\ -9.6& 3.6 /\ -2.7  \\
 B3LYP+D3$^{***}$         & 13.6 /\ 11.4 /\ 11.4&15.3 /\ 13.2  &22.2 /\  19.8 /\  -19.8& 6.4 /\ -6.1  &15.1 /\  12.5 /\ -11.6& 4.1 /\ -3.2 \\
\hline\hline
\end{tabular}
}
\end{center}
\begin{flushleft}
$^{*}$Pyrazole and s-Triazine are excluded\\
$^{**}$Computed from the explicitly thermally expanded structures obtained with the FIT force field\cite{jnum}\\
$^{***}$ Based on a small set of k-points (see Table S1)\\
\end{flushleft}
\end{table} 

\end{widetext}
%
\begin{widetext}
\begin{center}
\begin{table} [h!]
\caption{\label{tab:dftqm} Lattice energy (kJ/mol) and cell volume ($\AA^3$)  errors of X23 crystal set computed by  pure QM methods with respect to embedded QM:QM.}
\begin{tabular}{lcccc}
\hline\hline
 method &lattice energy error &lattice energy \% error &cell volume error &cell volume \% error \\
 (dispersion)       &(RMS /\ mean) & (RMS  /\ mean)&(RMS /\  mean) & (RMS /\  mean)\\
\hline
 PBE+D3 & 2.4/\ 1.4 &3.8/\ 2.1 &3.5/\ 2.1 & 0.9 /\ 0.6 \\
 PBE0+D3$^*$ & 4.7 /\ 4.1 & 5.5 /\ 5.0 & 10.4 /\ -9.5 &  2.8 /\ -2.8\\
\hline
 PBE+MBD\cite{gad} &  3.3/\ 2.8  & 5.0 /\ 4.0 &  6.0 /\ 4.8 &  1.7 /\ 1.4 \\
 PBE0+MBD$^*$ & 2.8 /\ 2.5 & 3.8 /\ 3.3 &  5.6/\ -4.6  &  1.5/\ -1.3 \\ 
\hline
 BLYP+D3 & 3.8 /\ -2.7 & 4.1 /\ -3.0 & 4.2 /\ 3.2 & 1.5 /\ 1.1 \\
 B3LYP+D3$^*$ & 1.7 /\ -0.2 &  1.6 /\ -0.4 & 2.7 /\ -2.0   & 0.7 /\ -0.6 \\
\hline\hline
\end{tabular}

\begin{flushleft}
$^*$Based on reduced set of k-points (see Table S1)\\
\end{flushleft}
\end{table} 
\end{center}
\end{widetext}

\section{\sffamily \Large Conclusions}
We propose a method for optimization of weakly bound periodic systems based on an embedding QM:QM scheme. In this approach, we combine the energies and gradients from high-level QM fragment calculations with those coming from fast, low-level periodic DFT. Here, we introduce PBE0 embedded into PBE and B3LYP embedded into BLYP as alternative method for  hybrid functionals for molecular crystals. The robust calculation of crystal lattice energies and cell volumes makes it the promising tool for further applications, especially when accurate structures and energetics from hybrid functionals are needed.

\section*{SUPPLEMENTARY MATERIAL}
See supplementary material for detailed lattice energies, cell volumes, reference cell volumes, and k-points. 
\begin{acknowledgments}
The authors gratefully acknowledge the support from Vienna Scientific Cluster (VSC-3).

\end{acknowledgments}

\bibliographystyle{apsrev}

\begin{thebibliography}{106}
\expandafter\ifx\csname natexlab\endcsname\relax\def\natexlab#1{#1}\fi
\expandafter\ifx\csname bibnamefont\endcsname\relax
  \def\bibnamefont#1{#1}\fi
\expandafter\ifx\csname bibfnamefont\endcsname\relax
  \def\bibfnamefont#1{#1}\fi
\expandafter\ifx\csname citenamefont\endcsname\relax
  \def\citenamefont#1{#1}\fi
\expandafter\ifx\csname url\endcsname\relax
  \def\url#1{\texttt{#1}}\fi
\expandafter\ifx\csname urlprefix\endcsname\relax\def\urlprefix{URL }\fi
\providecommand{\bibinfo}[2]{#2}
\providecommand{\eprint}[2][]{\url{#2}}

\bibitem[{\citenamefont{Reilly et~al.}(2016)\citenamefont{Reilly, Cooper,
  Adjiman, Bhattacharya, Boese, Brandenburg, Bygrave, Bylsma, Campbell, Car
  et~al.}}]{amreilly}
\bibinfo{author}{\bibfnamefont{A.~M.} \bibnamefont{Reilly}},
  \bibinfo{author}{\bibfnamefont{R.~I.} \bibnamefont{Cooper}},
  \bibinfo{author}{\bibfnamefont{C.~S.} \bibnamefont{Adjiman}},
  \bibinfo{author}{\bibfnamefont{S.}~\bibnamefont{Bhattacharya}},
  \bibinfo{author}{\bibfnamefont{A.~D.} \bibnamefont{Boese}},
  \bibinfo{author}{\bibfnamefont{J.~G.} \bibnamefont{Brandenburg}},
  \bibinfo{author}{\bibfnamefont{P.~J.} \bibnamefont{Bygrave}},
  \bibinfo{author}{\bibfnamefont{R.}~\bibnamefont{Bylsma}},
  \bibinfo{author}{\bibfnamefont{J.~E.} \bibnamefont{Campbell}},
  \bibinfo{author}{\bibfnamefont{R.}~\bibnamefont{Car}}, \bibnamefont{et~al.},
  \bibinfo{journal}{Acta Crystallogr. B} \textbf{\bibinfo{volume}{72}},
  \bibinfo{pages}{439} (\bibinfo{year}{2016}).

\bibitem[{\citenamefont{Lommerse et~al.}(2000)\citenamefont{Lommerse,
  Motherwell, Ammon, Dunitz, Gavezzotti, Hofmann, Leusen, Mooij, Price,
  Schweizer et~al.}}]{ccdc}
\bibinfo{author}{\bibfnamefont{J.}~\bibnamefont{Lommerse}},
  \bibinfo{author}{\bibfnamefont{W.}~\bibnamefont{Motherwell}},
  \bibinfo{author}{\bibfnamefont{H.}~\bibnamefont{Ammon}},
  \bibinfo{author}{\bibfnamefont{J.}~\bibnamefont{Dunitz}},
  \bibinfo{author}{\bibfnamefont{A.}~\bibnamefont{Gavezzotti}},
  \bibinfo{author}{\bibfnamefont{D.}~\bibnamefont{Hofmann}},
  \bibinfo{author}{\bibfnamefont{F.}~\bibnamefont{Leusen}},
  \bibinfo{author}{\bibfnamefont{W.}~\bibnamefont{Mooij}},
  \bibinfo{author}{\bibfnamefont{S.}~\bibnamefont{Price}},
  \bibinfo{author}{\bibfnamefont{B.}~\bibnamefont{Schweizer}},
  \bibnamefont{et~al.}, \bibinfo{journal}{Acta Crystallogr. B}
  \textbf{\bibinfo{volume}{56}}, \bibinfo{pages}{697} (\bibinfo{year}{2000}).

\bibitem[{\citenamefont{Brandenburg and Grimme}(2014)}]{jgb85}
\bibinfo{author}{\bibfnamefont{J.~G.} \bibnamefont{Brandenburg}}
  \bibnamefont{and} \bibinfo{author}{\bibfnamefont{S.}~\bibnamefont{Grimme}},
  \bibinfo{journal}{J. Phys. Chem. Lett.} \textbf{\bibinfo{volume}{5}},
  \bibinfo{pages}{1785} (\bibinfo{year}{2014}).

\bibitem[{\citenamefont{Dolgonos et~al.}(2018)\citenamefont{Dolgonos, Loboda,
  and Boese}}]{gad}
\bibinfo{author}{\bibfnamefont{G.~A.} \bibnamefont{Dolgonos}},
  \bibinfo{author}{\bibfnamefont{O.~A.} \bibnamefont{Loboda}},
  \bibnamefont{and} \bibinfo{author}{\bibfnamefont{A.~D.} \bibnamefont{Boese}},
  \bibinfo{journal}{J. Phys. Chem. A} \textbf{\bibinfo{volume}{122}},
  \bibinfo{pages}{708} (\bibinfo{year}{2018}).

\bibitem[{\citenamefont{Mortazavi et~al.}(2018)\citenamefont{Mortazavi,
  Brandenburg, Maurer, and Tkatchenko}}]{mmort}
\bibinfo{author}{\bibfnamefont{M.}~\bibnamefont{Mortazavi}},
  \bibinfo{author}{\bibfnamefont{J.~G.} \bibnamefont{Brandenburg}},
  \bibinfo{author}{\bibfnamefont{R.~J.} \bibnamefont{Maurer}},
  \bibnamefont{and}
  \bibinfo{author}{\bibfnamefont{A.}~\bibnamefont{Tkatchenko}},
  \bibinfo{journal}{J. Phys. Chem. Lett.} \textbf{\bibinfo{volume}{9}},
  \bibinfo{pages}{399} (\bibinfo{year}{2018}).

\bibitem[{\citenamefont{Cutini et~al.}(2016)\citenamefont{Cutini, Civalleri,
  Corno, Orlando, Brandenburg, Maschio, and Ugliengo}}]{mcut}
\bibinfo{author}{\bibfnamefont{M.}~\bibnamefont{Cutini}},
  \bibinfo{author}{\bibfnamefont{B.}~\bibnamefont{Civalleri}},
  \bibinfo{author}{\bibfnamefont{M.}~\bibnamefont{Corno}},
  \bibinfo{author}{\bibfnamefont{R.}~\bibnamefont{Orlando}},
  \bibinfo{author}{\bibfnamefont{J.~G.} \bibnamefont{Brandenburg}},
  \bibinfo{author}{\bibfnamefont{L.}~\bibnamefont{Maschio}}, \bibnamefont{and}
  \bibinfo{author}{\bibfnamefont{P.}~\bibnamefont{Ugliengo}},
  \bibinfo{journal}{J. Chem. Theory Comput.} \textbf{\bibinfo{volume}{12}},
  \bibinfo{pages}{3340} (\bibinfo{year}{2016}).

\bibitem[{\citenamefont{Johnson}(2017)}]{erj}
\bibinfo{author}{\bibfnamefont{E.~R.} \bibnamefont{Johnson}},
  \emph{\bibinfo{title}{Non-covalent Interactions in Quantum Chemistry and
  Physics}} (\bibinfo{publisher}{Elsevier}, \bibinfo{year}{2017}), p.
  \bibinfo{pages}{169}.

\bibitem[{\citenamefont{Nyman et~al.}(2016)\citenamefont{Nyman, Pundyke, and
  Day}}]{jnum}
\bibinfo{author}{\bibfnamefont{J.}~\bibnamefont{Nyman}},
  \bibinfo{author}{\bibfnamefont{O.~S.} \bibnamefont{Pundyke}},
  \bibnamefont{and} \bibinfo{author}{\bibfnamefont{G.~M.} \bibnamefont{Day}},
  \bibinfo{journal}{Phys. Chem. Chem. Phys.} \textbf{\bibinfo{volume}{18}},
  \bibinfo{pages}{15828} (\bibinfo{year}{2016}).

\bibitem[{\citenamefont{Otero-de-la Roza and Johnson}(2012)}]{aot}
\bibinfo{author}{\bibfnamefont{A.}~\bibnamefont{Otero-de-la Roza}}
  \bibnamefont{and} \bibinfo{author}{\bibfnamefont{E.~R.}
  \bibnamefont{Johnson}}, \bibinfo{journal}{J. Chem. Phys.}
  \textbf{\bibinfo{volume}{137}}, \bibinfo{pages}{054103}
  (\bibinfo{year}{2012}).

\bibitem[{\citenamefont{Perdew et~al.}(1996)\citenamefont{Perdew, Burke, and
  Ernzerhof}}]{pbe}
\bibinfo{author}{\bibfnamefont{J.~P.} \bibnamefont{Perdew}},
  \bibinfo{author}{\bibfnamefont{K.}~\bibnamefont{Burke}}, \bibnamefont{and}
  \bibinfo{author}{\bibfnamefont{M.}~\bibnamefont{Ernzerhof}},
  \bibinfo{journal}{Phys. Rev. Lett.} \textbf{\bibinfo{volume}{77}},
  \bibinfo{pages}{3865} (\bibinfo{year}{1996}).

\bibitem[{\citenamefont{Grimme et~al.}(2010)\citenamefont{Grimme, Antony,
  Ehrlich, and Krieg}}]{dftd3}
\bibinfo{author}{\bibfnamefont{S.}~\bibnamefont{Grimme}},
  \bibinfo{author}{\bibfnamefont{J.}~\bibnamefont{Antony}},
  \bibinfo{author}{\bibfnamefont{S.}~\bibnamefont{Ehrlich}}, \bibnamefont{and}
  \bibinfo{author}{\bibfnamefont{H.}~\bibnamefont{Krieg}}, \bibinfo{journal}{J.
  Chem. Phys.} \textbf{\bibinfo{volume}{132}}, \bibinfo{pages}{154104}
  (\bibinfo{year}{2010}).

\bibitem[{\citenamefont{Tkatchenko et~al.}(2012)\citenamefont{Tkatchenko,
  DiStasio, Car, and Scheffler}}]{mbd}
\bibinfo{author}{\bibfnamefont{A.}~\bibnamefont{Tkatchenko}},
  \bibinfo{author}{\bibfnamefont{R.~A.} \bibnamefont{DiStasio}},
  \bibinfo{author}{\bibfnamefont{R.}~\bibnamefont{Car}}, \bibnamefont{and}
  \bibinfo{author}{\bibfnamefont{M.}~\bibnamefont{Scheffler}},
  \bibinfo{journal}{Phys. Rev. Lett.} \textbf{\bibinfo{volume}{108}},
  \bibinfo{pages}{236402} (\bibinfo{year}{2012}).

\bibitem[{\citenamefont{Ambrosetti et~al.}(2014)\citenamefont{Ambrosetti,
  Reilly, and Tkatchenko}}]{mbdI}
\bibinfo{author}{\bibfnamefont{A.}~\bibnamefont{Ambrosetti}},
  \bibinfo{author}{\bibfnamefont{R.~A.} \bibnamefont{Reilly},
  \bibfnamefont{A.~M.~DiStasio}}, \bibnamefont{and}
  \bibinfo{author}{\bibfnamefont{A.}~\bibnamefont{Tkatchenko}},
  \bibinfo{journal}{J. Chem. Phys.} \textbf{\bibinfo{volume}{140}},
  \bibinfo{pages}{18A508} (\bibinfo{year}{2014}).

\bibitem[{\citenamefont{Kresse~et al.}(2016)}]{vasp}
\bibinfo{author}{\bibfnamefont{G.}~\bibnamefont{Kresse~et al.}},
  \bibinfo{journal}{VASP 5.4.1 http://www.vasp.at}  (\bibinfo{year}{2016}).

\bibitem[{\citenamefont{Clark et~al.}(2005)\citenamefont{Clark, Segall,
  Pickard, Hasnip, Probert, Refson, and Payne}}]{castep}
\bibinfo{author}{\bibfnamefont{S.~J.} \bibnamefont{Clark}},
  \bibinfo{author}{\bibfnamefont{M.~D.} \bibnamefont{Segall}},
  \bibinfo{author}{\bibfnamefont{C.~J.} \bibnamefont{Pickard}},
  \bibinfo{author}{\bibfnamefont{P.~J.} \bibnamefont{Hasnip}},
  \bibinfo{author}{\bibfnamefont{M.~J.} \bibnamefont{Probert}},
  \bibinfo{author}{\bibfnamefont{K.}~\bibnamefont{Refson}}, \bibnamefont{and}
  \bibinfo{author}{\bibfnamefont{M.~C.} \bibnamefont{Payne}},
  \bibinfo{journal}{Z. Kristallogr.} \textbf{\bibinfo{volume}{220}},
  \bibinfo{pages}{567} (\bibinfo{year}{2005}).

\bibitem[{\citenamefont{Giannozzi et~al.}(2009)\citenamefont{Giannozzi, Baroni,
  Bonini, Calandra, Car, Cavazzoni, Ceresoli, Chiarotti, Cococcioni, Dabo
  et~al.}}]{quanex}
\bibinfo{author}{\bibfnamefont{P.}~\bibnamefont{Giannozzi}},
  \bibinfo{author}{\bibfnamefont{S.}~\bibnamefont{Baroni}},
  \bibinfo{author}{\bibfnamefont{N.}~\bibnamefont{Bonini}},
  \bibinfo{author}{\bibfnamefont{M.}~\bibnamefont{Calandra}},
  \bibinfo{author}{\bibfnamefont{R.}~\bibnamefont{Car}},
  \bibinfo{author}{\bibfnamefont{C.}~\bibnamefont{Cavazzoni}},
  \bibinfo{author}{\bibfnamefont{D.}~\bibnamefont{Ceresoli}},
  \bibinfo{author}{\bibfnamefont{G.~L.} \bibnamefont{Chiarotti}},
  \bibinfo{author}{\bibfnamefont{M.}~\bibnamefont{Cococcioni}},
  \bibinfo{author}{\bibfnamefont{I.}~\bibnamefont{Dabo}}, \bibnamefont{et~al.},
  \bibinfo{journal}{J. Phys. Condens. Matter} \textbf{\bibinfo{volume}{21}},
  \bibinfo{pages}{395502} (\bibinfo{year}{2009}).

\bibitem[{\citenamefont{Hutter et~al.}(2014)\citenamefont{Hutter, Iannuzzi,
  Schiffmann, and Vandevondele}}]{cp2k}
\bibinfo{author}{\bibfnamefont{J.}~\bibnamefont{Hutter}},
  \bibinfo{author}{\bibfnamefont{M.}~\bibnamefont{Iannuzzi}},
  \bibinfo{author}{\bibfnamefont{F.}~\bibnamefont{Schiffmann}},
  \bibnamefont{and}
  \bibinfo{author}{\bibfnamefont{J.}~\bibnamefont{Vandevondele}},
  \bibinfo{journal}{Wiley Interdiscip. Rev. Comput. Mol. Sci.}
  \textbf{\bibinfo{volume}{4}}, \bibinfo{pages}{15} (\bibinfo{year}{2014}).

\bibitem[{\citenamefont{Boese}(2015)}]{bosecppc}
\bibinfo{author}{\bibfnamefont{A.~D.} \bibnamefont{Boese}},
  \bibinfo{journal}{ChemPhysChem} \textbf{\bibinfo{volume}{16}},
  \bibinfo{pages}{978} (\bibinfo{year}{2015}).

\bibitem[{\citenamefont{Peintinger et~al.}(2013)\citenamefont{Peintinger,
  Oliveira, and Bredow}}]{pient}
\bibinfo{author}{\bibfnamefont{M.~F.} \bibnamefont{Peintinger}},
  \bibinfo{author}{\bibfnamefont{D.~V.} \bibnamefont{Oliveira}},
  \bibnamefont{and} \bibinfo{author}{\bibfnamefont{T.}~\bibnamefont{Bredow}},
  \bibinfo{journal}{J. Comput. Chem.} \textbf{\bibinfo{volume}{34}},
  \bibinfo{pages}{451} (\bibinfo{year}{2013}).

\bibitem[{\citenamefont{Boese et~al.}(2007)\citenamefont{Boese, Martin, and
  Klopper}}]{klopper}
\bibinfo{author}{\bibfnamefont{A.~D.} \bibnamefont{Boese}},
  \bibinfo{author}{\bibfnamefont{J.~M.~L.} \bibnamefont{Martin}},
  \bibnamefont{and} \bibinfo{author}{\bibfnamefont{W.}~\bibnamefont{Klopper}},
  \bibinfo{journal}{J. Phys. Chem. A} \textbf{\bibinfo{volume}{111}},
  \bibinfo{pages}{11122} (\bibinfo{year}{2007}).

\bibitem[{\citenamefont{Boese et~al.}(2003)\citenamefont{Boese, Martin, and
  Handy}}]{mhandy}
\bibinfo{author}{\bibfnamefont{A.~D.} \bibnamefont{Boese}},
  \bibinfo{author}{\bibfnamefont{J.~M.~L.} \bibnamefont{Martin}},
  \bibnamefont{and} \bibinfo{author}{\bibfnamefont{N.~C.} \bibnamefont{Handy}},
  \bibinfo{journal}{J. Chem. Phys.} \textbf{\bibinfo{volume}{119}},
  \bibinfo{pages}{3005} (\bibinfo{year}{2003}).

\bibitem[{\citenamefont{Reilly and Tkatchenko}(2013)}]{amr}
\bibinfo{author}{\bibfnamefont{A.~M.} \bibnamefont{Reilly}} \bibnamefont{and}
  \bibinfo{author}{\bibfnamefont{A.}~\bibnamefont{Tkatchenko}},
  \bibinfo{journal}{J. Chem. Phys.} \textbf{\bibinfo{volume}{139}},
  \bibinfo{pages}{024705} (\bibinfo{year}{2013}).

\bibitem[{\citenamefont{Marom et~al.}(2013)\citenamefont{Marom, DiStasio,
  Atalla, Levchenko, Reilly, Chelikowsky, Leiserowitz, and Tkatchenko}}]{marom}
\bibinfo{author}{\bibfnamefont{N.}~\bibnamefont{Marom}},
  \bibinfo{author}{\bibfnamefont{R.~A.} \bibnamefont{DiStasio},
  \bibfnamefont{Jr.}},
  \bibinfo{author}{\bibfnamefont{V.}~\bibnamefont{Atalla}},
  \bibinfo{author}{\bibfnamefont{S.}~\bibnamefont{Levchenko}},
  \bibinfo{author}{\bibfnamefont{A.~M.} \bibnamefont{Reilly}},
  \bibinfo{author}{\bibfnamefont{J.~R.} \bibnamefont{Chelikowsky}},
  \bibinfo{author}{\bibfnamefont{L.}~\bibnamefont{Leiserowitz}},
  \bibnamefont{and}
  \bibinfo{author}{\bibfnamefont{A.}~\bibnamefont{Tkatchenko}},
  \bibinfo{journal}{Angew. Chem. Int. Ed.} \textbf{\bibinfo{volume}{52}},
  \bibinfo{pages}{6629} (\bibinfo{year}{2013}).

\bibitem[{\citenamefont{Shtukenberg et~al.}(2017)\citenamefont{Shtukenberg,
  Zhu, Carter, Vogt, Hoja, Schneider, Song, Pokroy, Polishchuk, Tkatchenko
  et~al.}}]{schtuk}
\bibinfo{author}{\bibfnamefont{A.~G.} \bibnamefont{Shtukenberg}},
  \bibinfo{author}{\bibfnamefont{Q.}~\bibnamefont{Zhu}},
  \bibinfo{author}{\bibfnamefont{D.~J.} \bibnamefont{Carter}},
  \bibinfo{author}{\bibfnamefont{L.}~\bibnamefont{Vogt}},
  \bibinfo{author}{\bibfnamefont{J.}~\bibnamefont{Hoja}},
  \bibinfo{author}{\bibfnamefont{E.}~\bibnamefont{Schneider}},
  \bibinfo{author}{\bibfnamefont{H.}~\bibnamefont{Song}},
  \bibinfo{author}{\bibfnamefont{B.}~\bibnamefont{Pokroy}},
  \bibinfo{author}{\bibfnamefont{I.}~\bibnamefont{Polishchuk}},
  \bibinfo{author}{\bibfnamefont{A.}~\bibnamefont{Tkatchenko}},
  \bibnamefont{et~al.}, \bibinfo{journal}{Chem. Sci.}
  \textbf{\bibinfo{volume}{8}}, \bibinfo{pages}{4926} (\bibinfo{year}{2017}).

\bibitem[{\citenamefont{Boese et~al.}(2013)\citenamefont{Boese, Kirchner,
  Echeverria, and Boese}}]{etylacet}
\bibinfo{author}{\bibfnamefont{A.~D.} \bibnamefont{Boese}},
  \bibinfo{author}{\bibfnamefont{M.}~\bibnamefont{Kirchner}},
  \bibinfo{author}{\bibfnamefont{G.~A.} \bibnamefont{Echeverria}},
  \bibnamefont{and} \bibinfo{author}{\bibfnamefont{R.}~\bibnamefont{Boese}},
  \bibinfo{journal}{ChemPhysChem} \textbf{\bibinfo{volume}{14}},
  \bibinfo{pages}{799} (\bibinfo{year}{2013}).

\bibitem[{\citenamefont{Sameera and Pantazis}(2012)}]{pantazis}
\bibinfo{author}{\bibfnamefont{W.~M.~C.} \bibnamefont{Sameera}}
  \bibnamefont{and} \bibinfo{author}{\bibfnamefont{D.~A.}
  \bibnamefont{Pantazis}}, \bibinfo{journal}{J. Chem. Theory Comput.}
  \textbf{\bibinfo{volume}{8}}, \bibinfo{pages}{2630} (\bibinfo{year}{2012}).

\bibitem[{\citenamefont{Kesharwani et~al.}(2016)\citenamefont{Kesharwani,
  Karton, and Martin}}]{manoj}
\bibinfo{author}{\bibfnamefont{M.~K.} \bibnamefont{Kesharwani}},
  \bibinfo{author}{\bibfnamefont{A.}~\bibnamefont{Karton}}, \bibnamefont{and}
  \bibinfo{author}{\bibfnamefont{J.~M.~L.} \bibnamefont{Martin}},
  \bibinfo{journal}{J. Chem. Theory Comput.} \textbf{\bibinfo{volume}{12}},
  \bibinfo{pages}{444} (\bibinfo{year}{2016}).

\bibitem[{\citenamefont{Karton}(2017)}]{karton}
\bibinfo{author}{\bibfnamefont{A.}~\bibnamefont{Karton}}, \bibinfo{journal}{J.
  Comp. Chem.} \textbf{\bibinfo{volume}{38}}, \bibinfo{pages}{370}
  (\bibinfo{year}{2017}).

\bibitem[{\citenamefont{\v{R}ez\'a\v{c}
  et~al.}(2018)\citenamefont{\v{R}ez\'a\v{c}, B\'im, Gutten, and
  Rul\'i\v{s}ek}}]{rezac}
\bibinfo{author}{\bibfnamefont{J.}~\bibnamefont{\v{R}ez\'a\v{c}}},
  \bibinfo{author}{\bibfnamefont{D.}~\bibnamefont{B\'im}},
  \bibinfo{author}{\bibfnamefont{O.}~\bibnamefont{Gutten}}, \bibnamefont{and}
  \bibinfo{author}{\bibfnamefont{L.}~\bibnamefont{Rul\'i\v{s}ek}},
  \bibinfo{journal}{J. Chem. Theory Comput.} \textbf{\bibinfo{volume}{14}},
  \bibinfo{pages}{1254} (\bibinfo{year}{2018}).

\bibitem[{\citenamefont{Gruzman et~al.}(2009)\citenamefont{Gruzman, Karton, and
  Martin}}]{gruzman}
\bibinfo{author}{\bibfnamefont{D.}~\bibnamefont{Gruzman}},
  \bibinfo{author}{\bibfnamefont{A.}~\bibnamefont{Karton}}, \bibnamefont{and}
  \bibinfo{author}{\bibfnamefont{J.~M.~L.} \bibnamefont{Martin}},
  \bibinfo{journal}{J. Phys. Chem. A} \textbf{\bibinfo{volume}{113}},
  \bibinfo{pages}{11974} (\bibinfo{year}{2009}).

\bibitem[{\citenamefont{Maschio et~al.}(2010)\citenamefont{Maschio, Usvyat,
  Sch{\"u}tz, and Civalleri}}]{masch}
\bibinfo{author}{\bibfnamefont{L.}~\bibnamefont{Maschio}},
  \bibinfo{author}{\bibfnamefont{D.}~\bibnamefont{Usvyat}},
  \bibinfo{author}{\bibfnamefont{M.}~\bibnamefont{Sch{\"u}tz}},
  \bibnamefont{and}
  \bibinfo{author}{\bibfnamefont{B.}~\bibnamefont{Civalleri}},
  \bibinfo{journal}{J. Chem. Phys.} \textbf{\bibinfo{volume}{132}},
  \bibinfo{pages}{134706} (\bibinfo{year}{2010}).

\bibitem[{\citenamefont{Del~Ben et~al.}(2012)\citenamefont{Del~Ben, Hutter, and
  VandeVondele}}]{delbenm}
\bibinfo{author}{\bibfnamefont{M.}~\bibnamefont{Del~Ben}},
  \bibinfo{author}{\bibfnamefont{J.}~\bibnamefont{Hutter}}, \bibnamefont{and}
  \bibinfo{author}{\bibfnamefont{J.}~\bibnamefont{VandeVondele}},
  \bibinfo{journal}{J. Chem. Theory Comput.} \textbf{\bibinfo{volume}{8}},
  \bibinfo{pages}{4177} (\bibinfo{year}{2012}).

\bibitem[{\citenamefont{Booth et~al.}(2012)\citenamefont{Booth, Gr{\"u}neis,
  Kresse, and Alavi}}]{booth}
\bibinfo{author}{\bibfnamefont{G.~H.} \bibnamefont{Booth}},
  \bibinfo{author}{\bibfnamefont{A.}~\bibnamefont{Gr{\"u}neis}},
  \bibinfo{author}{\bibfnamefont{G.}~\bibnamefont{Kresse}}, \bibnamefont{and}
  \bibinfo{author}{\bibfnamefont{A.}~\bibnamefont{Alavi}},
  \bibinfo{journal}{Nature} \textbf{\bibinfo{volume}{493}},
  \bibinfo{pages}{365} (\bibinfo{year}{2012}).

\bibitem[{\citenamefont{Suhai and Ladik}(1982)}]{suhai}
\bibinfo{author}{\bibfnamefont{S.}~\bibnamefont{Suhai}} \bibnamefont{and}
  \bibinfo{author}{\bibfnamefont{J.}~\bibnamefont{Ladik}}, \bibinfo{journal}{J.
  Phys. C: Solid State Phys.} \textbf{\bibinfo{volume}{15}},
  \bibinfo{pages}{4327−4337} (\bibinfo{year}{1982}).

\bibitem[{\citenamefont{Wen and Beran}(2011)}]{beran}
\bibinfo{author}{\bibfnamefont{S.}~\bibnamefont{Wen}} \bibnamefont{and}
  \bibinfo{author}{\bibfnamefont{G.}~\bibnamefont{Beran}}, \bibinfo{journal}{J.
  Chem. Theory Comput.} \textbf{\bibinfo{volume}{7}}, \bibinfo{pages}{3733}
  (\bibinfo{year}{2011}).

\bibitem[{\citenamefont{Nanda and Beran}(2012)}]{beranmp2}
\bibinfo{author}{\bibfnamefont{K.~D.} \bibnamefont{Nanda}} \bibnamefont{and}
  \bibinfo{author}{\bibfnamefont{G.~J.~O.} \bibnamefont{Beran}},
  \bibinfo{journal}{J. Chem. Phys.} \textbf{\bibinfo{volume}{137}},
  \bibinfo{pages}{174106} (\bibinfo{year}{2012}).

\bibitem[{\citenamefont{M{\"o}rschel and Schmidt}(2015)}]{schmidt}
\bibinfo{author}{\bibfnamefont{P.}~\bibnamefont{M{\"o}rschel}}
  \bibnamefont{and} \bibinfo{author}{\bibfnamefont{M.~U.}
  \bibnamefont{Schmidt}}, \bibinfo{journal}{Acta Crystallogr. A.}
  \textbf{\bibinfo{volume}{71}}, \bibinfo{pages}{26} (\bibinfo{year}{2015}).

\bibitem[{\citenamefont{Beran and Nanda}(2010)}]{benan}
\bibinfo{author}{\bibfnamefont{G.}~\bibnamefont{Beran}} \bibnamefont{and}
  \bibinfo{author}{\bibfnamefont{K.}~\bibnamefont{Nanda}}, \bibinfo{journal}{J.
  Phys. Chem. Lett.} \textbf{\bibinfo{volume}{1}}, \bibinfo{pages}{3480}
  (\bibinfo{year}{2010}).

\bibitem[{\citenamefont{Fang et~al.}(2015)\citenamefont{Fang, Li, Gu, and
  Li}}]{fang}
\bibinfo{author}{\bibfnamefont{T.}~\bibnamefont{Fang}},
  \bibinfo{author}{\bibfnamefont{W.}~\bibnamefont{Li}},
  \bibinfo{author}{\bibfnamefont{F.}~\bibnamefont{Gu}}, \bibnamefont{and}
  \bibinfo{author}{\bibfnamefont{S.}~\bibnamefont{Li}}, \bibinfo{journal}{J.
  Chem. Theor. Comput.} \textbf{\bibinfo{volume}{11}}, \bibinfo{pages}{91}
  (\bibinfo{year}{2015}).

\bibitem[{\citenamefont{Nolan et~al.}(2010)\citenamefont{Nolan, Bygrave, Allan,
  and Manby}}]{nolan}
\bibinfo{author}{\bibfnamefont{S.~J.} \bibnamefont{Nolan}},
  \bibinfo{author}{\bibfnamefont{P.~J.} \bibnamefont{Bygrave}},
  \bibinfo{author}{\bibfnamefont{N.~L.} \bibnamefont{Allan}}, \bibnamefont{and}
  \bibinfo{author}{\bibfnamefont{F.~R.} \bibnamefont{Manby}},
  \bibinfo{journal}{J. Phys. Condens. Matter} \textbf{\bibinfo{volume}{22}},
  \bibinfo{pages}{074201} (\bibinfo{year}{2010}).

\bibitem[{\citenamefont{\v{C}ervinka et~al.}(2016)\citenamefont{\v{C}ervinka,
  Fulem, and R\.u\v{z}i\v{c}ka}}]{cervinka}
\bibinfo{author}{\bibfnamefont{C.}~\bibnamefont{\v{C}ervinka}},
  \bibinfo{author}{\bibfnamefont{M.}~\bibnamefont{Fulem}}, \bibnamefont{and}
  \bibinfo{author}{\bibfnamefont{K.}~\bibnamefont{R\.u\v{z}i\v{c}ka}},
  \bibinfo{journal}{J. Chem. Phys.} \textbf{\bibinfo{volume}{144}},
  \bibinfo{pages}{064505} (\bibinfo{year}{2016}).

\bibitem[{\citenamefont{Stoll}(1992)}]{stoll}
\bibinfo{author}{\bibfnamefont{H.}~\bibnamefont{Stoll}},
  \bibinfo{journal}{Phys. Rev. B} \textbf{\bibinfo{volume}{46}},
  \bibinfo{pages}{6700} (\bibinfo{year}{1992}).

\bibitem[{\citenamefont{Paulus}(2006)}]{paulus}
\bibinfo{author}{\bibfnamefont{B.}~\bibnamefont{Paulus}},
  \bibinfo{journal}{Phys. Rep.} \textbf{\bibinfo{volume}{428}},
  \bibinfo{pages}{1} (\bibinfo{year}{2006}).

\bibitem[{\citenamefont{Doll and Stoll}(1997)}]{dollstoll}
\bibinfo{author}{\bibfnamefont{K.}~\bibnamefont{Doll}} \bibnamefont{and}
  \bibinfo{author}{\bibfnamefont{H.}~\bibnamefont{Stoll}},
  \bibinfo{journal}{Phys. Rev. B} \textbf{\bibinfo{volume}{56}},
  \bibinfo{pages}{10121} (\bibinfo{year}{1997}).

\bibitem[{\citenamefont{Hermann and Schwerdtfeger}(2009)}]{aherm}
\bibinfo{author}{\bibfnamefont{A.}~\bibnamefont{Hermann}} \bibnamefont{and}
  \bibinfo{author}{\bibfnamefont{P.}~\bibnamefont{Schwerdtfeger}},
  \bibinfo{journal}{J. Chem. Phys.} \textbf{\bibinfo{volume}{131}},
  \bibinfo{pages}{244508} (\bibinfo{year}{2009}).

\bibitem[{\citenamefont{Rosciszewski et~al.}(1999)\citenamefont{Rosciszewski,
  Paulus, Fulde, and Stoll}}]{rosci}
\bibinfo{author}{\bibfnamefont{K.}~\bibnamefont{Rosciszewski}},
  \bibinfo{author}{\bibfnamefont{B.}~\bibnamefont{Paulus}},
  \bibinfo{author}{\bibfnamefont{P.}~\bibnamefont{Fulde}}, \bibnamefont{and}
  \bibinfo{author}{\bibfnamefont{H.}~\bibnamefont{Stoll}},
  \bibinfo{journal}{Phys. Rev. B} \textbf{\bibinfo{volume}{60}},
  \bibinfo{pages}{7905} (\bibinfo{year}{1999}).

\bibitem[{\citenamefont{Hermann and Schwerdtfeger}(2008)}]{herm}
\bibinfo{author}{\bibfnamefont{A.}~\bibnamefont{Hermann}} \bibnamefont{and}
  \bibinfo{author}{\bibfnamefont{P.}~\bibnamefont{Schwerdtfeger}},
  \bibinfo{journal}{Phys. Rev. Lett.} \textbf{\bibinfo{volume}{101}},
  \bibinfo{pages}{183005} (\bibinfo{year}{2008}).

\bibitem[{\citenamefont{Friedrich et~al.}(2011)\citenamefont{Friedrich, Perlt,
  Roatsch, Spickermann, and Kirchner}}]{jfried}
\bibinfo{author}{\bibfnamefont{J.}~\bibnamefont{Friedrich}},
  \bibinfo{author}{\bibfnamefont{E.}~\bibnamefont{Perlt}},
  \bibinfo{author}{\bibfnamefont{M.}~\bibnamefont{Roatsch}},
  \bibinfo{author}{\bibfnamefont{C.}~\bibnamefont{Spickermann}},
  \bibnamefont{and} \bibinfo{author}{\bibfnamefont{B.}~\bibnamefont{Kirchner}},
  \bibinfo{journal}{J. Chem. Theor. Comp.} \textbf{\bibinfo{volume}{7}},
  \bibinfo{pages}{843} (\bibinfo{year}{2011}).

\bibitem[{\citenamefont{Friedrich
  et~al.}(2007{\natexlab{a}})\citenamefont{Friedrich, Hanrath, and
  Dolg}}]{jfriedr}
\bibinfo{author}{\bibfnamefont{J.}~\bibnamefont{Friedrich}},
  \bibinfo{author}{\bibfnamefont{M.}~\bibnamefont{Hanrath}}, \bibnamefont{and}
  \bibinfo{author}{\bibfnamefont{M.}~\bibnamefont{Dolg}}, \bibinfo{journal}{J.
  Chem. Phys.} \textbf{\bibinfo{volume}{126}}, \bibinfo{pages}{154110}
  (\bibinfo{year}{2007}{\natexlab{a}}).

\bibitem[{\citenamefont{Friedrich
  et~al.}(2007{\natexlab{b}})\citenamefont{Friedrich, Hanrath, and
  Dolg}}]{hanrat}
\bibinfo{author}{\bibfnamefont{J.}~\bibnamefont{Friedrich}},
  \bibinfo{author}{\bibfnamefont{M.}~\bibnamefont{Hanrath}}, \bibnamefont{and}
  \bibinfo{author}{\bibfnamefont{M.}~\bibnamefont{Dolg}}, \bibinfo{journal}{J.
  Phys. Chem. A} \textbf{\bibinfo{volume}{111}}, \bibinfo{pages}{9830}
  (\bibinfo{year}{2007}{\natexlab{b}}).

\bibitem[{\citenamefont{M{\"u}ller and Paulus}(2012)}]{mull}
\bibinfo{author}{\bibfnamefont{C.}~\bibnamefont{M{\"u}ller}} \bibnamefont{and}
  \bibinfo{author}{\bibfnamefont{B.}~\bibnamefont{Paulus}},
  \bibinfo{journal}{Phys. Chem. Chem. Phys.} \textbf{\bibinfo{volume}{14}},
  \bibinfo{pages}{7605} (\bibinfo{year}{2012}).

\bibitem[{\citenamefont{Wen et~al.}(2012)\citenamefont{Wen, Nanda, Huang, and
  Beran}}]{beranqmmm}
\bibinfo{author}{\bibfnamefont{S.}~\bibnamefont{Wen}},
  \bibinfo{author}{\bibfnamefont{K.}~\bibnamefont{Nanda}},
  \bibinfo{author}{\bibfnamefont{Y.}~\bibnamefont{Huang}}, \bibnamefont{and}
  \bibinfo{author}{\bibfnamefont{G.~J.~O.} \bibnamefont{Beran}},
  \bibinfo{journal}{Phys. Chem. Chem. Phys.} \textbf{\bibinfo{volume}{14}},
  \bibinfo{pages}{7578} (\bibinfo{year}{2012}).

\bibitem[{\citenamefont{Boese and Sauer}(2017)}]{sauer}
\bibinfo{author}{\bibfnamefont{A.~D.} \bibnamefont{Boese}} \bibnamefont{and}
  \bibinfo{author}{\bibfnamefont{J.}~\bibnamefont{Sauer}},
  \bibinfo{journal}{Cryst. Growth Des.} \textbf{\bibinfo{volume}{17}},
  \bibinfo{pages}{1636} (\bibinfo{year}{2017}).

\bibitem[{\citenamefont{Becke}(1988)}]{becke}
\bibinfo{author}{\bibfnamefont{A.}~\bibnamefont{Becke}},
  \bibinfo{journal}{Phys. Rev. A} \textbf{\bibinfo{volume}{3}},
  \bibinfo{pages}{3098} (\bibinfo{year}{1988}).

\bibitem[{\citenamefont{Lee et~al.}(1998)\citenamefont{Lee, Yang, and
  Parr}}]{lyp}
\bibinfo{author}{\bibfnamefont{C.}~\bibnamefont{Lee}},
  \bibinfo{author}{\bibfnamefont{W.}~\bibnamefont{Yang}}, \bibnamefont{and}
  \bibinfo{author}{\bibfnamefont{R.}~\bibnamefont{Parr}},
  \bibinfo{journal}{Phys. Rev. B: Condens. Matter Mater. Phys.}
  \textbf{\bibinfo{volume}{37}}, \bibinfo{pages}{785} (\bibinfo{year}{1998}).

\bibitem[{\citenamefont{Heyd et~al.}(2003)\citenamefont{Heyd, Scuseria, and
  Ernzerhof}}]{hse03}
\bibinfo{author}{\bibfnamefont{J.}~\bibnamefont{Heyd}},
  \bibinfo{author}{\bibfnamefont{G.~E.} \bibnamefont{Scuseria}},
  \bibnamefont{and}
  \bibinfo{author}{\bibfnamefont{M.}~\bibnamefont{Ernzerhof}},
  \bibinfo{journal}{J. Chem. Phys.} \textbf{\bibinfo{volume}{118}},
  \bibinfo{pages}{8207} (\bibinfo{year}{2003}).

\bibitem[{\citenamefont{Krukau et~al.}(2006)\citenamefont{Krukau, Vydrov,
  Izmaylov, and Scuseria}}]{hse06}
\bibinfo{author}{\bibfnamefont{A.~V.} \bibnamefont{Krukau}},
  \bibinfo{author}{\bibfnamefont{O.~A.} \bibnamefont{Vydrov}},
  \bibinfo{author}{\bibfnamefont{A.~F.} \bibnamefont{Izmaylov}},
  \bibnamefont{and} \bibinfo{author}{\bibfnamefont{G.~E.}
  \bibnamefont{Scuseria}}, \bibinfo{journal}{J. Chem. Phys.}
  \textbf{\bibinfo{volume}{125}}, \bibinfo{pages}{224106}
  (\bibinfo{year}{2006}).

\bibitem[{\citenamefont{Gaus et~al.}(2011)\citenamefont{Gaus, Cui, and
  Elstner}}]{dftb3}
\bibinfo{author}{\bibfnamefont{M.}~\bibnamefont{Gaus}},
  \bibinfo{author}{\bibfnamefont{Q.}~\bibnamefont{Cui}}, \bibnamefont{and}
  \bibinfo{author}{\bibfnamefont{M.}~\bibnamefont{Elstner}},
  \bibinfo{journal}{J. Chem. Theor. Comp.} \textbf{\bibinfo{volume}{7}},
  \bibinfo{pages}{931} (\bibinfo{year}{2011}).

\bibitem[{\citenamefont{Tkatchenko and Scheffler}(2009)}]{dftts}
\bibinfo{author}{\bibfnamefont{A.}~\bibnamefont{Tkatchenko}} \bibnamefont{and}
  \bibinfo{author}{\bibfnamefont{M.}~\bibnamefont{Scheffler}},
  \bibinfo{journal}{Phys. Rev. Lett.} \textbf{\bibinfo{volume}{102}},
  \bibinfo{pages}{073005} (\bibinfo{year}{2009}).

\bibitem[{\citenamefont{Sure and Grimme}(2013)}]{hf3c}
\bibinfo{author}{\bibfnamefont{R.}~\bibnamefont{Sure}} \bibnamefont{and}
  \bibinfo{author}{\bibfnamefont{S.}~\bibnamefont{Grimme}},
  \bibinfo{journal}{J. Comput. Chem.} \textbf{\bibinfo{volume}{34}},
  \bibinfo{pages}{1672} (\bibinfo{year}{2013}).

\bibitem[{\citenamefont{Grimme}(2006)}]{dftd2}
\bibinfo{author}{\bibfnamefont{S.}~\bibnamefont{Grimme}}, \bibinfo{journal}{J.
  Comput. Chem.} \textbf{\bibinfo{volume}{27}}, \bibinfo{pages}{1787}
  (\bibinfo{year}{2006}).

\bibitem[{\citenamefont{Moellman and Grimme}(2014)}]{jmoel}
\bibinfo{author}{\bibfnamefont{J.}~\bibnamefont{Moellman}} \bibnamefont{and}
  \bibinfo{author}{\bibfnamefont{S.}~\bibnamefont{Grimme}},
  \bibinfo{journal}{J. Phys. Chem. C} \textbf{\bibinfo{volume}{118}},
  \bibinfo{pages}{7615} (\bibinfo{year}{2014}).

\bibitem[{\citenamefont{Hammer et~al.}(1999)\citenamefont{Hammer, Hansen, and
  N{\o}rskov}}]{rpbe}
\bibinfo{author}{\bibfnamefont{B.}~\bibnamefont{Hammer}},
  \bibinfo{author}{\bibfnamefont{L.~B.} \bibnamefont{Hansen}},
  \bibnamefont{and} \bibinfo{author}{\bibfnamefont{J.~K.}
  \bibnamefont{N{\o}rskov}}, \bibinfo{journal}{Phys. Rev. B}
  \textbf{\bibinfo{volume}{59}}, \bibinfo{pages}{7413} (\bibinfo{year}{1999}).

\bibitem[{\citenamefont{Sato and Nakai}(2009)}]{lrd}
\bibinfo{author}{\bibfnamefont{T.}~\bibnamefont{Sato}} \bibnamefont{and}
  \bibinfo{author}{\bibfnamefont{H.}~\bibnamefont{Nakai}}, \bibinfo{journal}{J.
  Chem. Phys.} \textbf{\bibinfo{volume}{131}}, \bibinfo{pages}{224104}
  (\bibinfo{year}{2009}).

\bibitem[{\citenamefont{Ikabata et~al.}(2015)\citenamefont{Ikabata, Tsukamoto,
  Imamura, and Nakai}}]{yik}
\bibinfo{author}{\bibfnamefont{Y.}~\bibnamefont{Ikabata}},
  \bibinfo{author}{\bibfnamefont{Y.}~\bibnamefont{Tsukamoto}},
  \bibinfo{author}{\bibfnamefont{Y.}~\bibnamefont{Imamura}}, \bibnamefont{and}
  \bibinfo{author}{\bibfnamefont{H.}~\bibnamefont{Nakai}}, \bibinfo{journal}{J.
  Comput. Chem.} \textbf{\bibinfo{volume}{36}}, \bibinfo{pages}{303}
  (\bibinfo{year}{2015}).

\bibitem[{\citenamefont{Zhang and Yang}(1998)}]{revpbe}
\bibinfo{author}{\bibfnamefont{Y.}~\bibnamefont{Zhang}} \bibnamefont{and}
  \bibinfo{author}{\bibfnamefont{W.}~\bibnamefont{Yang}},
  \bibinfo{journal}{Phys. Rev. Lett.} \textbf{\bibinfo{volume}{80}},
  \bibinfo{pages}{890} (\bibinfo{year}{1998}).

\bibitem[{\citenamefont{Gould et~al.}(2016)\citenamefont{Gould, Leb\`egue,
  \'Angy\'an, and Bu\v{c}ko}}]{tgoul}
\bibinfo{author}{\bibfnamefont{T.}~\bibnamefont{Gould}},
  \bibinfo{author}{\bibfnamefont{S.}~\bibnamefont{Leb\`egue}},
  \bibinfo{author}{\bibfnamefont{J.~G.} \bibnamefont{\'Angy\'an}},
  \bibnamefont{and}
  \bibinfo{author}{\bibfnamefont{T.}~\bibnamefont{Bu\v{c}ko}},
  \bibinfo{journal}{J. Chem. Theory Comput.} \textbf{\bibinfo{volume}{12}},
  \bibinfo{pages}{5920} (\bibinfo{year}{2016}).

\bibitem[{\citenamefont{Bu\v{c}ko et~al.}(2016)\citenamefont{Bu\v{c}ko,
  Leb\`egue, Gould, and \'Angy\'an}}]{tbuc}
\bibinfo{author}{\bibfnamefont{T.}~\bibnamefont{Bu\v{c}ko}},
  \bibinfo{author}{\bibfnamefont{S.}~\bibnamefont{Leb\`egue}},
  \bibinfo{author}{\bibfnamefont{T.}~\bibnamefont{Gould}}, \bibnamefont{and}
  \bibinfo{author}{\bibfnamefont{J.~G.} \bibnamefont{\'Angy\'an}},
  \bibinfo{journal}{J. Phys. Condens. Matter} \textbf{\bibinfo{volume}{28}},
  \bibinfo{pages}{045201} (\bibinfo{year}{2016}).

\bibitem[{\citenamefont{Becke and Johnson}(2005)}]{bj}
\bibinfo{author}{\bibfnamefont{A.~D.} \bibnamefont{Becke}} \bibnamefont{and}
  \bibinfo{author}{\bibfnamefont{E.~R.} \bibnamefont{Johnson}},
  \bibinfo{journal}{J. Chem. Phys.} \textbf{\bibinfo{volume}{123}},
  \bibinfo{pages}{154101} (\bibinfo{year}{2005}).

\bibitem[{\citenamefont{Johnson and Becke}(2006)}]{bjI}
\bibinfo{author}{\bibfnamefont{E.}~\bibnamefont{Johnson}} \bibnamefont{and}
  \bibinfo{author}{\bibfnamefont{A.}~\bibnamefont{Becke}}, \bibinfo{journal}{J.
  Chem. Phys.} \textbf{\bibinfo{volume}{124}}, \bibinfo{pages}{174104}
  (\bibinfo{year}{2006}).

\bibitem[{\citenamefont{Johnson and Becke}(2007)}]{bjIII}
\bibinfo{author}{\bibfnamefont{E.~R.} \bibnamefont{Johnson}} \bibnamefont{and}
  \bibinfo{author}{\bibfnamefont{A.~D.} \bibnamefont{Becke}},
  \bibinfo{journal}{J. Chem. Phys.} \textbf{\bibinfo{volume}{127}},
  \bibinfo{pages}{154108} (\bibinfo{year}{2007}).

\bibitem[{\citenamefont{Johnson}()}]{erjh}
\bibinfo{author}{\bibfnamefont{E.~R.} \bibnamefont{Johnson}},
  \bibinfo{howpublished}{personal communication}.

\bibitem[{\citenamefont{Perdew}(1986)}]{pw86}
\bibinfo{author}{\bibfnamefont{J.~P.} \bibnamefont{Perdew}},
  \bibinfo{journal}{Phys. Rev. B} \textbf{\bibinfo{volume}{33}},
  \bibinfo{pages}{8822} (\bibinfo{year}{1986}).

\bibitem[{\citenamefont{Becke}(1986)}]{b86b}
\bibinfo{author}{\bibfnamefont{A.~D.} \bibnamefont{Becke}},
  \bibinfo{journal}{J. Chem. Phys.} \textbf{\bibinfo{volume}{85}},
  \bibinfo{pages}{7184} (\bibinfo{year}{1986}).

\bibitem[{\citenamefont{Thonhauser et~al.}(2007)\citenamefont{Thonhauser,
  Cooper, Li, Puzder, Hyldgaard, and Langreth}}]{optB88}
\bibinfo{author}{\bibfnamefont{T.}~\bibnamefont{Thonhauser}},
  \bibinfo{author}{\bibfnamefont{V.~R.} \bibnamefont{Cooper}},
  \bibinfo{author}{\bibfnamefont{S.}~\bibnamefont{Li}},
  \bibinfo{author}{\bibfnamefont{A.}~\bibnamefont{Puzder}},
  \bibinfo{author}{\bibfnamefont{P.}~\bibnamefont{Hyldgaard}},
  \bibnamefont{and} \bibinfo{author}{\bibfnamefont{D.~C.}
  \bibnamefont{Langreth}}, \bibinfo{journal}{Phys. Rev. B}
  \textbf{\bibinfo{volume}{76}}, \bibinfo{pages}{125112}
  (\bibinfo{year}{2007}).

\bibitem[{\citenamefont{Klime\v{s} et~al.}(2010)\citenamefont{Klime\v{s},
  Bowler, and Michaelides}}]{df2}
\bibinfo{author}{\bibfnamefont{J.}~\bibnamefont{Klime\v{s}}},
  \bibinfo{author}{\bibfnamefont{D.~R.} \bibnamefont{Bowler}},
  \bibnamefont{and}
  \bibinfo{author}{\bibfnamefont{A.}~\bibnamefont{Michaelides}},
  \bibinfo{journal}{J. Phys. Condens. Matter} \textbf{\bibinfo{volume}{22}},
  \bibinfo{pages}{022201} (\bibinfo{year}{2010}).

\bibitem[{\citenamefont{Tao et~al.}(2003)\citenamefont{Tao, Perdew, Staroverov,
  and Scuseria}}]{tpss}
\bibinfo{author}{\bibfnamefont{J.}~\bibnamefont{Tao}},
  \bibinfo{author}{\bibfnamefont{J.}~\bibnamefont{Perdew}},
  \bibinfo{author}{\bibfnamefont{V.}~\bibnamefont{Staroverov}},
  \bibnamefont{and} \bibinfo{author}{\bibfnamefont{G.}~\bibnamefont{Scuseria}},
  \bibinfo{journal}{Phys. Rev. Lett.} \textbf{\bibinfo{volume}{91}},
  \bibinfo{pages}{146401} (\bibinfo{year}{2003}).

\bibitem[{\citenamefont{Bl{\"o}chl}(1994)}]{paw}
\bibinfo{author}{\bibfnamefont{P.~E.} \bibnamefont{Bl{\"o}chl}},
  \bibinfo{journal}{Phys. Rev. B} \textbf{\bibinfo{volume}{50}},
  \bibinfo{pages}{17953} (\bibinfo{year}{1994}).

\bibitem[{\citenamefont{Grimme et~al.}(2015)\citenamefont{Grimme, Brandenburg,
  Bannwarth, and Hansen}}]{sgr}
\bibinfo{author}{\bibfnamefont{S.}~\bibnamefont{Grimme}},
  \bibinfo{author}{\bibfnamefont{J.~G.} \bibnamefont{Brandenburg}},
  \bibinfo{author}{\bibfnamefont{C.}~\bibnamefont{Bannwarth}},
  \bibnamefont{and} \bibinfo{author}{\bibfnamefont{A.}~\bibnamefont{Hansen}},
  \bibinfo{journal}{J. Chem. Phys.} \textbf{\bibinfo{volume}{143}},
  \bibinfo{pages}{054107} (\bibinfo{year}{2015}).

\bibitem[{\citenamefont{Becke}(1993)}]{b3lypI}
\bibinfo{author}{\bibfnamefont{A.}~\bibnamefont{Becke}}, \bibinfo{journal}{J.
  Phys. Chem.} \textbf{\bibinfo{volume}{98}}, \bibinfo{pages}{5648}
  (\bibinfo{year}{1993}).

\bibitem[{\citenamefont{Stephens et~al.}(1994)\citenamefont{Stephens, Devlin,
  Chabalowski, and Frisch}}]{b3lypII}
\bibinfo{author}{\bibfnamefont{P.}~\bibnamefont{Stephens}},
  \bibinfo{author}{\bibfnamefont{F.}~\bibnamefont{Devlin}},
  \bibinfo{author}{\bibfnamefont{C.}~\bibnamefont{Chabalowski}},
  \bibnamefont{and} \bibinfo{author}{\bibfnamefont{M.}~\bibnamefont{Frisch}},
  \bibinfo{journal}{J. Phys. Chem.} \textbf{\bibinfo{volume}{98}},
  \bibinfo{pages}{11623} (\bibinfo{year}{1994}).

\bibitem[{\citenamefont{Civalleri et~al.}(2008)\citenamefont{Civalleri,
  Zicovich-Wilson, Valenzano, and Ugliengo}}]{civa}
\bibinfo{author}{\bibfnamefont{B.}~\bibnamefont{Civalleri}},
  \bibinfo{author}{\bibfnamefont{C.}~\bibnamefont{Zicovich-Wilson}},
  \bibinfo{author}{\bibfnamefont{L.}~\bibnamefont{Valenzano}},
  \bibnamefont{and} \bibinfo{author}{\bibfnamefont{P.}~\bibnamefont{Ugliengo}},
  \bibinfo{journal}{Cryst. Eng. Comm.} \textbf{\bibinfo{volume}{10}},
  \bibinfo{pages}{405} (\bibinfo{year}{2008}).

\bibitem[{\citenamefont{Sch{\"a}fer et~al.}(1992)\citenamefont{Sch{\"a}fer,
  Horn, and Ahlrichs}}]{tzp}
\bibinfo{author}{\bibfnamefont{A.}~\bibnamefont{Sch{\"a}fer}},
  \bibinfo{author}{\bibfnamefont{H.}~\bibnamefont{Horn}}, \bibnamefont{and}
  \bibinfo{author}{\bibfnamefont{R.}~\bibnamefont{Ahlrichs}},
  \bibinfo{journal}{J. Chem. Phys.} \textbf{\bibinfo{volume}{97}},
  \bibinfo{pages}{2571} (\bibinfo{year}{1992}).

\bibitem[{\citenamefont{Adamo and Barone}(1999)}]{pbe0}
\bibinfo{author}{\bibfnamefont{C.}~\bibnamefont{Adamo}} \bibnamefont{and}
  \bibinfo{author}{\bibfnamefont{V.}~\bibnamefont{Barone}},
  \bibinfo{journal}{J. Chem. Phys.} \textbf{\bibinfo{volume}{110}},
  \bibinfo{pages}{6158} (\bibinfo{year}{1999}).

\bibitem[{\citenamefont{Johnson and Becke}(2005)}]{bj0}
\bibinfo{author}{\bibfnamefont{E.~R.} \bibnamefont{Johnson}} \bibnamefont{and}
  \bibinfo{author}{\bibfnamefont{A.~D.} \bibnamefont{Becke}},
  \bibinfo{journal}{J. Chem. Phys.} \textbf{\bibinfo{volume}{123}},
  \bibinfo{pages}{024101} (\bibinfo{year}{2005}).

\bibitem[{\citenamefont{Grimme et~al.}(2011)\citenamefont{Grimme, Ehrlich, and
  Goerigk}}]{bjII}
\bibinfo{author}{\bibfnamefont{S.}~\bibnamefont{Grimme}},
  \bibinfo{author}{\bibfnamefont{S.}~\bibnamefont{Ehrlich}}, \bibnamefont{and}
  \bibinfo{author}{\bibfnamefont{L.}~\bibnamefont{Goerigk}},
  \bibinfo{journal}{J. Comput. Chem.} \textbf{\bibinfo{volume}{32}},
  \bibinfo{pages}{1456} (\bibinfo{year}{2011}).

\bibitem[{tur(2015)}]{turbo}
\emph{\bibinfo{title}{{TURBOMOLE V7.0}, a development of {University of
  Karlsruhe} and {Forschungszentrum Karlsruhe GmbH}, 1989-2007, {TURBOMOLE
  GmbH}, since 2007; available from {\tt http://www.turbomole.com}}}
  (\bibinfo{year}{2015}).

\bibitem[{\citenamefont{Rappoport and Furche}(2010)}]{def2}
\bibinfo{author}{\bibfnamefont{D.}~\bibnamefont{Rappoport}} \bibnamefont{and}
  \bibinfo{author}{\bibfnamefont{F.}~\bibnamefont{Furche}},
  \bibinfo{journal}{J. Chem. Phys.} \textbf{\bibinfo{volume}{133}},
  \bibinfo{pages}{134105} (\bibinfo{year}{2010}).

\bibitem[{\citenamefont{Gaus et~al.}(2013)\citenamefont{Gaus, Goez, and
  Elstner}}]{dftbIII}
\bibinfo{author}{\bibfnamefont{M.}~\bibnamefont{Gaus}},
  \bibinfo{author}{\bibfnamefont{A.}~\bibnamefont{Goez}}, \bibnamefont{and}
  \bibinfo{author}{\bibfnamefont{M.}~\bibnamefont{Elstner}},
  \bibinfo{journal}{J. Chem. Theory Comput.} \textbf{\bibinfo{volume}{9}},
  \bibinfo{pages}{338} (\bibinfo{year}{2013}).

\bibitem[{\citenamefont{Aradi et~al.}(2007)\citenamefont{Aradi, Hourahine, and
  Frauenheim}}]{dftbI}
\bibinfo{author}{\bibfnamefont{B.}~\bibnamefont{Aradi}},
  \bibinfo{author}{\bibfnamefont{B.}~\bibnamefont{Hourahine}},
  \bibnamefont{and}
  \bibinfo{author}{\bibfnamefont{T.}~\bibnamefont{Frauenheim}},
  \bibinfo{journal}{J. Phys. Chem. A} \textbf{\bibinfo{volume}{111}},
  \bibinfo{pages}{5678} (\bibinfo{year}{2007}).

\bibitem[{dft(accessed May 3, 2017)}]{dftbII}
\bibinfo{journal}{DFTB+ version 1.3.1.}  (\bibinfo{year}{accessed May 3,
  2017}), \urlprefix\url{http://www.dftb-plus.info}.

\bibitem[{\citenamefont{Sierka et~al.}(2011)\citenamefont{Sierka, Tuma, and
  Kerber}}]{qmpot}
\bibinfo{author}{\bibfnamefont{M.}~\bibnamefont{Sierka}},
  \bibinfo{author}{\bibfnamefont{C.}~\bibnamefont{Tuma}}, \bibnamefont{and}
  \bibinfo{author}{\bibfnamefont{T.}~\bibnamefont{Kerber}},
  \bibinfo{journal}{Humboldt Universit{\"a}t, Berlin}  (\bibinfo{year}{2011}).

\bibitem[{\citenamefont{Bu\v{c}ko et~al.}(2005)\citenamefont{Bu\v{c}ko, Hafner,
  and \'Angy\'an}}]{angy}
\bibinfo{author}{\bibfnamefont{T.}~\bibnamefont{Bu\v{c}ko}},
  \bibinfo{author}{\bibfnamefont{J.}~\bibnamefont{Hafner}}, \bibnamefont{and}
  \bibinfo{author}{\bibfnamefont{J.~G.} \bibnamefont{\'Angy\'an}},
  \bibinfo{journal}{J. Chem. Phys.} \textbf{\bibinfo{volume}{122}},
  \bibinfo{pages}{124508} (\bibinfo{year}{2005}).

\bibitem[{\citenamefont{Doll}(2010)}]{doll}
\bibinfo{author}{\bibfnamefont{K.}~\bibnamefont{Doll}}, \bibinfo{journal}{Mol.
  Phys.} \textbf{\bibinfo{volume}{108}}, \bibinfo{pages}{223}
  (\bibinfo{year}{2010}).

\bibitem[{\citenamefont{Knuth et~al.}(2015)\citenamefont{Knuth, Carbogno,
  Atalla, Blum, and Scheffler}}]{knuth}
\bibinfo{author}{\bibfnamefont{F.}~\bibnamefont{Knuth}},
  \bibinfo{author}{\bibfnamefont{C.}~\bibnamefont{Carbogno}},
  \bibinfo{author}{\bibfnamefont{V.}~\bibnamefont{Atalla}},
  \bibinfo{author}{\bibfnamefont{V.}~\bibnamefont{Blum}}, \bibnamefont{and}
  \bibinfo{author}{\bibfnamefont{M.}~\bibnamefont{Scheffler}},
  \bibinfo{journal}{Comput. Phys. Commun.} \textbf{\bibinfo{volume}{190}},
  \bibinfo{pages}{33} (\bibinfo{year}{2015}).

\bibitem[{\citenamefont{Press et~al.}(1996)\citenamefont{Press, Teukolsky,
  Vetterling, and Flannery}}]{numrecipes}
\bibinfo{author}{\bibfnamefont{W.~H.} \bibnamefont{Press}},
  \bibinfo{author}{\bibfnamefont{S.~A.} \bibnamefont{Teukolsky}},
  \bibinfo{author}{\bibfnamefont{W.~T.} \bibnamefont{Vetterling}},
  \bibnamefont{and} \bibinfo{author}{\bibfnamefont{B.~P.}
  \bibnamefont{Flannery}}, \emph{\bibinfo{title}{Numerical Recipes in Fortran
  90 (2nd ed.): the Art of Parallel Scientific Computing.}}
  (\bibinfo{publisher}{Cambridge University Press}, \bibinfo{address}{New
  York}, \bibinfo{year}{1996}).

\bibitem[{\citenamefont{Francis and Payne}(1990)}]{gpf}
\bibinfo{author}{\bibfnamefont{G.~P.} \bibnamefont{Francis}} \bibnamefont{and}
  \bibinfo{author}{\bibfnamefont{M.~C.} \bibnamefont{Payne}},
  \bibinfo{journal}{J. Phys.: Condens. Matter} \textbf{\bibinfo{volume}{2}},
  \bibinfo{pages}{4395} (\bibinfo{year}{1990}).

\bibitem[{\citenamefont{Heit et~al.}(2016)\citenamefont{Heit, Nanda, and
  Beran}}]{beranco}
\bibinfo{author}{\bibfnamefont{Y.~N.} \bibnamefont{Heit}},
  \bibinfo{author}{\bibfnamefont{K.~D.} \bibnamefont{Nanda}}, \bibnamefont{and}
  \bibinfo{author}{\bibfnamefont{G.~J.~O.} \bibnamefont{Beran}},
  \bibinfo{journal}{Chem. Sci.} \textbf{\bibinfo{volume}{7}},
  \bibinfo{pages}{246} (\bibinfo{year}{2016}).

\bibitem[{\citenamefont{Erba et~al.}(2016)\citenamefont{Erba, Maul, and
  Civalleri}}]{civall}
\bibinfo{author}{\bibfnamefont{A.}~\bibnamefont{Erba}},
  \bibinfo{author}{\bibfnamefont{J.}~\bibnamefont{Maul}}, \bibnamefont{and}
  \bibinfo{author}{\bibfnamefont{B.}~\bibnamefont{Civalleri}},
  \bibinfo{journal}{Chem. Commun.} \textbf{\bibinfo{volume}{52}},
  \bibinfo{pages}{1820} (\bibinfo{year}{2016}).

\bibitem[{\citenamefont{Becka and Cruickshank}(1963)}]{becka}
\bibinfo{author}{\bibfnamefont{L.~N.} \bibnamefont{Becka}} \bibnamefont{and}
  \bibinfo{author}{\bibfnamefont{D.~W.~J.} \bibnamefont{Cruickshank}},
  \bibinfo{journal}{Proc. R. Soc. Lond. A} \textbf{\bibinfo{volume}{273}},
  \bibinfo{pages}{435} (\bibinfo{year}{1963}).

\bibitem[{\citenamefont{Heit and Beran}(2016)}]{beranheit}
\bibinfo{author}{\bibfnamefont{Y.~N.} \bibnamefont{Heit}} \bibnamefont{and}
  \bibinfo{author}{\bibfnamefont{G.~J.~O.} \bibnamefont{Beran}},
  \bibinfo{journal}{Acta Cryst.} \textbf{\bibinfo{volume}{B72}},
  \bibinfo{pages}{514} (\bibinfo{year}{2016}).

\bibitem[{\citenamefont{Hoja et~al.}(2017)\citenamefont{Hoja, Reilly, and
  Tkatchenko}}]{hoja}
\bibinfo{author}{\bibfnamefont{J.}~\bibnamefont{Hoja}},
  \bibinfo{author}{\bibfnamefont{A.~M.} \bibnamefont{Reilly}},
  \bibnamefont{and}
  \bibinfo{author}{\bibfnamefont{A.}~\bibnamefont{Tkatchenko}},
  \bibinfo{journal}{WIREs Comput. Mol. Sci.} \textbf{\bibinfo{volume}{7}},
  \bibinfo{pages}{e1294} (\bibinfo{year}{2017}).

\bibitem[{\citenamefont{David et~al.}(1992)\citenamefont{David, Ibberson,
  Jeffrey, and Ruble}}]{rubl}
\bibinfo{author}{\bibfnamefont{W.}~\bibnamefont{David}},
  \bibinfo{author}{\bibfnamefont{R.}~\bibnamefont{Ibberson}},
  \bibinfo{author}{\bibfnamefont{G.}~\bibnamefont{Jeffrey}}, \bibnamefont{and}
  \bibinfo{author}{\bibfnamefont{J.}~\bibnamefont{Ruble}},
  \bibinfo{journal}{Physica B} \textbf{\bibinfo{volume}{180}},
  \bibinfo{pages}{597} (\bibinfo{year}{1992}).

\bibitem[{\citenamefont{Kabsch}(1976)}]{kabsch}
\bibinfo{author}{\bibfnamefont{W.}~\bibnamefont{Kabsch}},
  \bibinfo{journal}{Acta Crystallogr. A} \textbf{\bibinfo{volume}{32}},
  \bibinfo{pages}{922} (\bibinfo{year}{1976}).

\bibitem[{\citenamefont{Kabsch}(1978)}]{kabsch2}
\bibinfo{author}{\bibfnamefont{W.}~\bibnamefont{Kabsch}},
  \bibinfo{journal}{Acta Crystallogr. A} \textbf{\bibinfo{volume}{34}},
  \bibinfo{pages}{827} (\bibinfo{year}{1978}).

\bibitem[{\citenamefont{Temelso et~al.}(2017)\citenamefont{Temelso, Mabey,
  Kubota, Appiah-Padi, and Shields}}]{arbalign}
\bibinfo{author}{\bibfnamefont{B.}~\bibnamefont{Temelso}},
  \bibinfo{author}{\bibfnamefont{J.~M.} \bibnamefont{Mabey}},
  \bibinfo{author}{\bibfnamefont{T.}~\bibnamefont{Kubota}},
  \bibinfo{author}{\bibfnamefont{N.}~\bibnamefont{Appiah-Padi}},
  \bibnamefont{and} \bibinfo{author}{\bibfnamefont{G.~C.}
  \bibnamefont{Shields}}, \bibinfo{journal}{J. Chem. Inf. Model.}
  \textbf{\bibinfo{volume}{57}}, \bibinfo{pages}{1045} (\bibinfo{year}{2017}).

\end{thebibliography}
%

\end{document}


\onecolumngrid
\begin{center}
{\large
Supplementary Material \\
\vspace{0.5cm}
{\bf Towards Hybrid Density Functional Calculations of Molecular Crystals via Fragment-Based Methods}\\
\vspace{0.5cm}
O. A. Loboda$^{1,2,a}$, G. A. Dolgonos$^{1}$, A. D. Boese$^{1,b}$}\\
\end{center}

 $^{1}$:  Institute of Chemistry, University of Graz, Heinrichstrasse 28/IV, A-8010 Graz, Austria\\
 $^{2}$: On leave from : A.V. Dumansky Institute of Colloid and Water Chemistry, National Academy of Sciences of Ukraine, Vernadsky bld. 42, Kyiv-142, Ukraine \\

$^a$: E-mail: oleksandr.loboda@uni-graz.at \\
$^b$: http://www.chemie.uni-graz.at/en/quantum-chemistry \\
\begin{center}
{\bf Contents}\\
\end{center}
\hspace*{11cm} Page\\
General information\hspace{8.2cm} 2\\
Table S1\hspace{10cm} 2\\
Table S2\hspace{10cm} 3\\
Table S3\hspace{10cm} 3\\
Table S4\hspace{10cm} 3\\
Table S5\hspace{10cm} 4\\
Table S6\hspace{10cm} 4\\
Table S7\hspace{10cm} 4\\
Table S8\hspace{10cm} 5\\
Table S9\hspace{10cm} 5\\
Table S10\hspace{9.8cm} 6\\
Figure S1 \hspace{9.7cm} 6\\
\null\clearpage
\begin{center}
{\large General information }\\ 
\end{center}

Experimental data are taken from \citet{mmort}
The quality of a model is measured by quantifying the mean (ME) and root mean square errors (RMS) of lattice energies and volumes:
\begin{align}
ME=\frac{1}{N}\sum_{i=1}^{N}( x^{model}- x^{exp})_i=\frac{1}{23}\sum_{i=1}^{23}\delta_{i}
\end{align}

\begin{align}
\label{Eq2}
RMS  = \sqrt{(\frac{1}{23}) \sum_{i=1}^{23}\delta_{i}^2}
\end{align}
The respective ME \% and RMS \% errors are calculated similarly based on 
\begin{align}
\delta_{i}(\%)  = \frac{( x^{model}- x^{exp})_i\times100\%}{x^{exp}_i}
\end{align}

\begin{center}
\begin{table} [h!]
\caption*{Table S1. \label{mean} Mean absolute errors (MAE) of lattice energies $E$ (kJ/mol) of X23 molecular crystals  with respect to experiment calculated  without lattice relaxation and reported in the literature. For a comparison, some representative fully-lattice-optimized MAE($E$) values from Table I are listed in a last column.
 }
\begin{tabular}{lcccc}
\hline\hline
Method &  \multicolumn{4}{c}{MAE($E$)} \\
       &          &\multicolumn{2}{c}{Using fully optimized geometry}& \\
       &          & PBE+TS & TPSS+D3 & lattice opt. \\   
\hline
MSINDO-D3H+ & 27.6\cite{jgb75}& & & \\
DFTB3+D3 & 9.86$^c$\cite{jnum}, 10.0$^c$\cite{jgb75}, 10.4$^c$\cite{jgb85} & & &13.3$^a$\cite{gad} \\
HF-3c & 6.3.$^{b,c}$\cite{jgb45}, 7.1$^{c}$\cite{jgb45}, 9.2$^{c}$\cite{jgb75}, 13.0$^{c,d}$\cite{jgb75} & & & 8.2\cite{mcut}  \\
PBE+D3 & 3.9$^c$\cite{mcut}, 4.48$^c$\cite{jgb85,jgb45}, 5.06$^{b,c}$\cite{jgb45} & & & 5.5$^a$\cite{gad} \\
PBE+MBD & 6.40$^c$\cite{jgb45}  & 5.91\cite{amr} & & 6.2$^a$\cite{gad} \\
PBE+TS & 6.40$^c$\cite{jgb45}& & & 15.7$^a$\cite{gad} \\
PBE+XDM & 6.28$^c$\cite{jgb45}& & & 6.51\cite{jnum,aot} \\
B866b+XDM & 5.73$^c$\cite{jgb45}& & & \\
PBE0+TS & &10.02\cite{amr}& &\\
PBE0+MBD & &3.92\cite{amr}& &\\
PBE0+D3 & & &5.0\cite{jmoel},5.0$^{b}$\cite{jmoel}& \\
HSE06+D3 & & & 4.2$^{b}$\cite{jmoel}, 5.0\cite{jmoel} & \\
\hline\hline 
\end{tabular}

\begin{flushleft}
$^a$Calculated based on data from Supporting Information.\\
$^b$With the three-body contribution included. \\
$^c$Keeping frozen the experimental values of lattice parameters.\\
$^d$Using effective core potentials for core electrons.\\
\end{flushleft}
\end{table} 
\end{center}

\begin{center}
\begin{table} [h!]
\caption*{Table S2. Grids of \label{mean}k-points used in periodic DFT calculations}
\begin{tabular}{cccc}
\hline\hline
X23 phase              &  small    &   standard         & large    \\
\hline
 Acetic acid           &  1 4 3    &    2 5 4           & 3 7 6     \\
 Adamantane            &    3 3 2  &    3 3 2           & 5 5 4     \\
 Anthracene            &    2 3 2  &    3 4 2           & 7 8 6      \\
 Benzene               &    2 2 2  &    3 2 3           & 4 3 5     \\
 Carbon dioxide        &  4 4 4    &    4 4 4           & 6 6 6     \\
 Cyanamide             &    2 2 2  &    3 3 2           & 4 4 3     \\
 Cyclohexane-1,4-dione &  2 2 2    &    3 3 3           & 6 6 6     \\
 Cytosine              &    1 2 4  &    2 2 6           & 4 4 10    \\
 Ethyl carbamate       &    3 2 2  &    4 3 3           & 6 5 5     \\
 Formamide             &    4 2 2  &    6 2 3           & 9 4 5     \\
 Hexamethylenetetramine&    3 3 3  &    3 3 3           & 5 5 5     \\
 Imidazole             &    2 3 2  &    3 4 3           & 4 6 4     \\
 Naphthalene           &    2 3 2  &    3 4 3           & 4 5 4     \\
 Ammonia               &   4 4 4   &    4 4 4           & 6 6 6     \\
 Oxalic acid ($\alpha$)&    2 2 2  &    3 3 4           & 5 4 5     \\
 Oxalic acid ($\beta$) &    3 2 3  &    4 3 4           & 6 5 6     \\
 Pyrazine              &    2 3 4  &    2 4 6           & 3 5 8     \\
 Pyrazole              &    2 1 2  &    4 3 2           & 6 5 4     \\
 Succinic acid         &    3 2 3  &    4 3 4           & 6 4 6     \\
 s-Triazine            &  2 2 2    &    2 3 3           & 4 6 6     \\
 s-Trioxane            &    2 2 2  &    2 3 3           & 3 4 4     \\
 Uracil                &  1 1 4    &    2 2 6           & 4 4 8     \\
 Urea                  &    4 4 5  &    4 4 5           & 6 6 7     \\
\hline\hline
\end{tabular}
\end{table}
\end{center}

\begin{table}[!]
\begin {center}
\caption*{Table S3.\label{qmqm} Methane binding energies to the empty hydrate (kJ/mol). All calculations were performed with either TURBOMOLE (TM) or VASP program package.}
\hspace*{-1.9cm}\begin{tabular}{lcc}
\hline
Hi / Lo&PBE+D3 (TM)&PBE+D3(VASP)\\
\hline
 PBE0+D3 (TM)&-14.14&-14.70 \\
 TPSS+D3 (TM)&-10.05&-11.57 \\
 MP2 (TM)& -19.40&-20.05\\  
\hline
\hline
\end{tabular}
\end{center}
\end{table}

\begin{center}
\begin{table} [h!]
\caption*{Table S4.\label{pbe0mbd} PBE0+MBD volumes ( \AA$^3$) for 10 structures obtained with small ans standard k-point grids and respective errors to experiment.}
\begin{tabular}{lccccc}
\hline\hline
                   &  \multicolumn{2}{c}{PBE0+MBD} &  Exp.\cite{mmort} &  \multicolumn{2}{c}{Error} \\
                   &  small    &   standard       &          & small      & standard    \\
\hline
Acetic acid        &   289.85  &  289.85          & 297.3    &  -7.45     &   -7.45\\
Benzene            &   449.55  &   449.49         & 474.1   &  -24.55    &   -24.61\\
Carbon dioxide     &   182.55  &   182.55$^{*}$         & 177.9   &    4.65    &     4.65\\
Cyanamide          &   413.25  &   412.86 (-0.09)$^{**}$& 415.7 &   -2.45    &    -2.84\\
Formamide          &   219.04  &   219.59 (+0.25)$^{**}$& 224.1 &   -5.06    &    -4.51\\
Naphthalene        &   332.90  &   332.90         & 340.8   &   -7.90     &    -7.90\\
Ammonia            &   121.30  &   121.30$^{*}$   & 135.1   &  -13.80     &   -13.80\\
Oxalic acid ($\alpha$)   &   303.87  &   303.86         & 312.6   &   -8.73    &    -8.74\\
Oxalic acid ($\beta$)   &   153.17  &   153.17         & 156.9   &   -3.73    &    -3.73\\
Pyrazine            &   195.54  &   195.55         & 203.6   &   -8.06    &    -8.05\\
\hline\hline
\end{tabular}
\begin{flushleft}
$^*$  Large k-point set was used\\
$^{**}$ Maximum and minimum deviations (to small k values) in \%.\\
\end{flushleft}
\end{table}
\end{center}

\begin{table}
\caption*{\label{tab:sp10} Table S5. Lattice volumes  ($\AA^3$) of the X23 molecular crystals with and without thermal correction.}

\hspace*{-2.5cm}
\scalebox{0.95}{\begin{tabular}{lccccccc}
\hline
\hline
X23 structure         &   Ref.  &FIT ff & \% deviation & W99rev6311P5 & \% deviation &  other &     \% deviation \\
\hline
Acetic acid           &  297.27	&292.43	&-1.63	&290.60	&-2.25	&290.73\cite{beranheit} 	&-2.2   \\
Adamantane            &  393.07	&378.78	&-3.64	&370.53	&-5.73	&		           &\\
Anthracene            &  456.47	&449.99	&-1.42	&448.15	&-1.82	&		           &\\
Benzene               &  474.07	&462.56	&-2.43	&426.56	&-10.02	&		         &\\
Carbon dioxide        &  177.88	&	&	&	&	&163.83\cite{beranco}; 163.47\cite{krupskii}  	&-7.9;  -8.1\\
Cyanamide             &  415.65	&426.30	&2.56	&396.06	&-4.71	&		           &\\
Cyclohexane-1,4-dione &  279.55	&272.75	&-2.43	&271.27	&-2.96	&		           &\\
Cytosine              &  472.42	&462.79	&-2.04	&452.20	&-4.28	&		           &\\
Ethyl carbamate       &  248.77	&242.02	&-2.72	&239.07	&-3.90	&		           &\\
Formamide             &  224.08	&217.79	&-2.81	&216.05	&-3.58	&		           &\\
Hexamethylenetetramine&  329.9	&340.88	&-1.80	&340.58	&-1.89	&          	           &\\
Imidazole             &  348.76	&339.43	&-2.68	&336.08	&-3.63	&337.60\cite{beranheit} 	&-3.2  \\
Naphthalene           &  340.83	&336.16	&-1.37	&334.99	&-1.71	&		           &\\
Ammonia               &  135.04	&127.22	&-5.79	&112.26	&-16.87	&128.02\cite{hoja} 	&-5.2  \\
Oxalic acid ($\alpha$)&  312.59	&296.50	&-5.15	&292.31	&-6.49  &                          &\\
Oxalic acid ($\beta$) &  156.87	&147.55	&-5.94	&142.61	&-9.09  &                          &\\
Pyrazine              &  203.64	&194.57	&-4.45	&192.32	&-5.56	&		           &\\
Pyrazole              &  698.27	&648.23	&-7.17	&635.19	&-9.03	&		           &\\
Succinic acid         &  243.9	&234.51	&-2.00	&231.68	&-3.18	&		           &\\
s-Triazine            &  586.82	&522.42	&-10.97	&521.02	&-11.21	&		           &\\
s-Trioxane            &  616.54	&601.24	&-2.48	&599.41	&-2.78	&		           &\\
Uracil                &  463.39	&443.36	&-4.32	&439.68	&-5.11	&		           &\\
Urea                  &  145.1	&	&	&	&	&143.5039\cite{civall}	&-1.1 \\
\hline
\hline
\hline
\end{tabular}}
\end{table}

\begin{center}
\begin{table} [h!]
\caption*{Table S6. Cell volume ($\AA^3$)  errors of complete X23 and truncated (X23-2)$^{*}$ crystal sets computed by embedded QM:QM and  pure QM methods with respect to experimental reference values without (Ref.) and with (FIT ff, W99) thermal expansion correction$^{**}$.}
\hspace*{-1.7cm}
\scalebox{0.75}{\begin{tabular}{lcccccccc}
\hline\hline
 method &  &  \%      &   & \%    &   & \% & & \% \\
 (dispersion)      & RMS /\ MAE   /\ mean  & RMS  /\ mean & RMS /\ MAE   /\ mean  & RMS  /\ mean  & RMS /\  MAE   /\ mean & RMS /\ mean   & RMS /\ MAE   /\ mean  & RMS  /\ mean \\
            &   \multicolumn{2}{c}{Ref. X23}  &   \multicolumn{2}{c}{Ref. X23-2}  &        \multicolumn{2}{c}{FIT ff. X23-2}     &  \multicolumn{2}{c}{W99. X23-2}                       \\ 
\hline
 PBE+D3                &9.3  /\  6.3  /\ -4.3  &2.9 /\ -1.2    &7.0  /\  5.2  /\ -3.6  &2.8 /\ -1.1    &9.3 /\  7.0  /\ 4.7  &4.1 /\  1.9         & 11.6 /\ 9.5  /\ 8.5  & 4.8 /\ 3.1 \\ 
 PBE0:PBE(PBE0+D3)     &10.1 /\  7.4  /\ -6.4  &3.1 /\ -1.8    &8.1 /\  6.4  /\ -5.3   &3.0 /\ -1.7    &7.5  /\ 5.6  /\ 2.9  &3.7 /\  1.3         & 10.1 /\ 7.9   /\ 6.8  & 4.4 /\ 2.5 \\
 PBE0+D3$^{***}$           &18.7 /\  16.0 /\ -15.9 &5.0 /\ -4.5    &16.2 /\  14.3 /\ -14.3 &4.9 /\ -4.4    &10.4 /\  8.4 /\ -6.0 &3.4 /\ -1.5         & 7.9 /\ 6.1   /\ -2.2 & 3.1 /\ -0.4\\
\hline
 PBE+MBD\cite{gad}     & 8.1 /\   5.5  /\  -2.1 &  2.9 /\ -0.7 & 7.1 /\   4.6  /\  -2.2 &  2.9 /\ -0.7 &10.0 /\   7.7  /\  6.1&  4.5 /\ 2.4       & 13.4 /\ 10.5  /\ 9.9  & 5.3 /\ 3.5 \\
 PBE0:PBE(PBE0+MBD)    &10.4 /\   7.7  /\  -6.9 &  3.2 /\ -2.0 &9.8 /\   7.3  /\  -6.4 &  3.3 /\ -2.0  &7.9 /\   6.3  /\  1.8 &  3.5 /\ 1.0       & 10.2 /\ 8.1   /\ 5.7  & 4.1 /\ 2.1 \\
 PBE0+MBD$^{***}$          &14.3 /\   12.0 /\ -11.6 &  4.1/\ -3.4  &13.1 /\   11.1 /\ -10.6 &  4.1/\ -3.3  &8.1 /\  6.5 /\ -2.4  &  3.4/\ -0.4        & 8.2 /\ 6.7   /\ 1.4 & 3.5 /\ 0.8 \\
\hline
 BLYP+D3               &21.3 /\  17.0 /\  -17.0& 5.3 /\ -4.6   &17.6 /\  14.8 /\ -14.8 & 5.0 /\ -4.5   &12.2 /\  9.7 /\  -6.5& 3.6 /\ -1.5        & 9.8 /\ 7.3  /\ -2.7 & 3.4 /\ -0.4\\
 B3LYP:BLYP(B3LYP+D3)  &23.6 /\  20.2 /\  -20.2& 6.0 /\ -5.7   &19.9 /\  17.8 /\  -17.8& 5.8 /\ -5.6   &13.0 /\  10.8 /\ -9.6& 3.6 /\ -2.7        & 9.9 /\ 8.0  /\ -5.7 & 3.0 /\ -1.6\\
 B3LYP+D3$^{***}$          &25.8 /\  22.2 /\  -22.2& 6.6 /\ -6.3   &22.2 /\  19.8 /\  -19.8& 6.4 /\ -6.1   &15.1 /\  12.5 /\ -11.6& 4.1 /\ -3.2       & 11.8 /\ 9.3 /\ -7.7& 3.3 /\ -2.1\\
\hline\hline
\end{tabular}
}
\begin{flushleft}
$^*$Pyrazole and s-Triazine are excluded\\
$^{**}$Computed from the explicitly thermally expanded structures obtained with the FIT and W99rev6311P5 force fields\cite{jnum}\\
$^{***}$Based on a small set of k-points (see Table S1)\\
\end{flushleft}
\end{table} 
\end{center}

\begin{center}
\begin{table} [h!]
\caption*{Table S7. Mean absolute errors (MAE) of cell volumes $V$ ($\AA^3$) of X23-2 molecular crystals calculated with different theoretical models reported in the literature with respect to  experimental reference values without (Ref.) and with (FIT ff) thermal expansion correction\cite{jnum,gad}}. 
\begin{tabular}{lcc}
\hline\hline
Method & \multicolumn{2}{c}{ MAE($V$)$^*$}\\
 & Ref. & FIT ff. \\
\hline
\hline
DFTB3\cite{dftb3}+D3\cite{dftd3} & 44.7 & 36.6\\
PBE\cite{pbe}+D3\cite{dftd3} & 5.2 & 6.9 \\
RPBE\cite{rpbe}+D3\cite{dftd3} & 9.0 & 13.4\\
PBE\cite{rpbe}+MBD\cite{mbd,mbdI} &  4.6 &7.6\\
PBE\cite{pbe}+TS\cite{dftts} &  7.9 &6.1\\
BLYP\cite{becke,lyp}+D3\cite{dftd3}& 14.8  & 9.6\\
BLYP\cite{becke,lyp}+D3\cite{dftd3}:DFTB3\cite{dftb3}+D3\cite{dftd3} &16.1 &  11.4\\
optB88-vdW\cite{optB88} & 12.8 & 5.6\\
vdW-DF2\cite{df2} & 6.9 &13.3\\
\hline\hline 
\end{tabular}
\begin{flushleft}
\hspace*{-0.52cm}$^*$Calculated based on data from Supporting Information.\\
\end{flushleft}
\end{table} 
\end{center}

\begin{table} [h!]
\caption*{Table S8. Calculated and experimental values of lattice energies (kJ/mol) for each X23 crystal phase with various dispersion-corrected DFT functionals and QM:QM. Dispersion term is given in parentheses.}
\scalebox{0.98}{\begin{tabular}{lcccccccccc}
\hline
\hline
{\multirow{2}{*}{ X23 structure}}&PBE           &PBE0:PBE       & PBE0  &PBE    &PBE0:PBE   &PBE0    & BLYP     &B3LYP:BLYP &B3LYP      & {\multirow{2}{*}{Exp.\cite{mmort}}}     \\
                                 & (D3)         &(PBE0+D3)      & (D3)  & (MBD) &(PBE0+MBD) & (MBD)  &  (D3)    &(B3LYP+D3) &  (D3)     &          \\
\hline
 Acetic acid                     & 74.52	&  72.78        &76.05	&76.18	&  73.48    &75.64   &74.34	&  79.22    & 78.47	&72.80  \\      
 Adamantane                      & 72.12	&  68.93        &72.30	&77.82	&  71.26    &75.49   &81.06	&  82.07    & 80.43	&69.40  \\
 Anthracene                      & 126.76	&  110.63  	&115.29	&104.57	&  104.85   &106.65  &130.48	&  126.76   & 127.06	&112.70  \\
 Benzene                         & 55.10	&  54.69        &59.75	&53.33	&  50.94    &54.52   &63.94	&  64.64    & 64.49	&55.30  \\
 Carbon dioxide                  & 24.44	&  24.33        &25.30	&23.05	&  22.41    &23.02   &26.79	&  29.70    & 29.33	&28.40  \\
 Cyanamide                       & 93.25	&  88.28        &91.95	&92.19	&  86.17    &88.9    &93.38	&  94.60    & 94.29	&79.70  \\
 Cyclohexane-1,4-dione           & 88.81	&  88.11        &96.81	&90.77	&  88.81    &93.42   &97.06	&  101.39   & 104.30	&88.60  \\
 Cytosine                        & 172.75	&  159.86  	&164.67	&159.45	&  157.04   &159.42  &169.63	&  172.77   & 172.52	&162.80  \\
 Ethyl carbamate                 & 87.95	&  86.17        &90.72	&90.21	&  87.58    &90.09   &90.38	&  94.13    & 95.10	&86.30  \\
 Formamide                       & 82.13	&  80.48        &83.05	&81.72	&  79.52    &81.21   &82.57	&  86.68    & 85.59	&79.20  \\
 Hexamethylenetetramine          & 86.17	&  85.38        &86.64	&88.62	&  86.22    &86.49   &93.69	&  96.36    & 93.61	&86.20  \\
 Imidazole                       & 92.03	&  89.75        &92.45	&92.24	&  88.87    &90.5    &94.96	&  96.78    & 95.33	&86.80  \\
 Naphthalene                     & 80.11	&  82.13        &86.72	&78.74	&  77.37    &79.56   &96.63	&  94.16    & 96.07	&83.10  \\
 Ammonia                         & 43.07	&  38.10        &41.07	&42.65	&  37.73    &40.45   &39.89	&  41.50    & 40.85	&37.20  \\
 Oxalic acid ($\alpha$)          & 127.18	&  116.19  	&121.70	&117.55	&  117.21   &120.25  &118.07	&  127.18   & 127.01	&96.30  \\
 Oxalic acid ($\beta$)           & 125.97	&  115.97  	&122.11	&119.92	&  116.56   &120.49  &119.87	&  126.23   & 126.74	&96.10  \\
 Pyrazine                        & 65.67	&  63.58        &67.09	&61.54	&  57.67    &60.13   &75.47	&  75.81    & 74.57	&61.30  \\
 Pyrazole                        & 81.24	&  78.79        &83.29	&81.27	&  77.94    &80.24   &84.40	&  85.81    & 86.11	&77.70  \\
 Succinic acid                   & 141.84	&  129.07  	&134.65	&136.38	&  131.37   &135.05  &135.77	&  141.93   & 141.05	&130.30  \\
 s-Triazine                      & 60.73	&  59.23        &62.83	&56.35	&  53.43    &55.78   &70.31	&  71.49    & 70.43	&61.70  \\
 s-Trioxane                      & 58.29	&  55.32        &59.07	&62.48	&  58.34    &60.93   &63.93	&  66.57    & 66.13	&66.40  \\
 Uracil                          & 149.28	&  137.19  	&146.76	&136.54	&  134.31   &139.54  &146.67	&  149.28   & 154.09	&135.70  \\
 Urea                            & 118.03	&  111.91  	&111.99	&112.95	&  112.21   &111.84  &112.62	&  118.03   & 115.25	&102.50  \\
\hline
\end{tabular}
}
\end{table}

\begin{table} [h!]
\caption*{Table S9. Calculated values of cell volumes (\AA$^3$) for each X23 crystal phase with various dispersion-corrected DFT functionals and QM:QM. Dispersion term is given in parentheses.}
\scalebox{0.98}{\begin{tabular}{lccccccccc}
\hline
\hline
{\multirow{2}{*}{ X23 structure}}&PBE    &PBE0:PBE & PBE0   &PBE    &PBE0:PBE   &PBE0    & BLYP   &B3LYP:BLYP &B3LYP    \\
                                 & (D3)  &(PBE0+D3)&   (D3) & (MBD) &(PBE0+MBD) & (MBD)  &  (D3)  &(B3LYP+D3) &  (D3)    \\
\hline
Acetic acid           &  297.59& 297.11   &  288.97& 297.31 & 296.53  &  289.85 & 290.36 & 286.01      & 283.66   \\ 
Adamantane            &  377.82& 376.84   &  367.00& 373.14 & 365.38  &  363.09 & 365.01 & 362.80      & 360.34   \\
Anthracene            &  449.90& 447.54   &  434.33& 455.28 & 449.02  &  443.06 & 429.37 & 429.44      & 426.46   \\
Benzene               &  457.02& 456.44   &  441.39& 460.48 & 458.41  &  449.55 & 435.07 & 434.69      & 430.40   \\
Carbon dioxide        &  186.38& 184.68   &  178.45& 188.94 & 184.39  &  182.55 & 175.79 & 170.56      & 169.10   \\
Cyanamide             &  413.63& 418.21   &  402.02& 422.77 & 418.45  &  413.25 & 400.46 & 399.89      & 396.69   \\
Cyclohexane-1,4-dione &  276.10& 272.85   &  266.37& 276.28 & 271.62  &  268.72 & 267.13 & 264.06      & 261.86   \\
Cytosine              &  466.49& 461.17   &  450.74& 473.68 & 469.51  &  459.26 & 449.32 & 443.92      & 444.19   \\
Ethyl carbamate       &  241.81& 239.39   &  233.30& 240.72 & 234.38  &  234.23 & 234.33 & 230.45      & 229.23   \\
Formamide             &  222.81& 220.92   &  215.34& 224.58 & 222.15  &  219.04 & 214.14 & 211.44      & 210.14   \\
Hexamethylenetetramine&  333.78& 330.54   &  322.04& 330.22 & 322.55  &  322.32 & 325.07 & 321.66      & 320.05   \\
Imidazole             &  345.89& 345.51   &  333.79& 347.28 & 344.50  &  336.77 & 330.16 & 328.33      & 326.87   \\
Naphthalene           &  337.79& 337.68   &  325.63& 340.94 & 340.85  &  332.90 & 322.26 & 322.24      & 319.71   \\
Ammonia               &  122.59& 122.74   &  120.85& 122.93 & 123.46  &  121.30 & 122.97 & 122.78      & 122.30   \\
Oxalic acid ($\alpha$)&  314.53& 308.12   &  300.72& 315.75 & 310.65  &  303.87 & 308.95 & 300.97      & 298.99   \\
Oxalic acid ($\beta$) &  158.70& 158.33   &  151.51& 158.99 & 153.49  &  153.17 & 153.75 & 149.08      & 148.91   \\
Pyrazine              &  196.38& 196.10   &  190.94& 200.04 & 199.42  &  195.54 & 186.62 & 186.28      & 185.43   \\
Pyrazole              &  704.26& 694.04   &  677.08& 711.38 & 695.25  &  687.57 & 674.77 & 665.13      & 665.22   \\
Succinic acid         &  244.90& 243.90   &  236.16& 244.10 & 238.99  &  236.90 & 236.72 & 230.77      & 231.10   \\
s-Triazine            &  556.47& 556.02   &  541.19& 570.15 & 565.27  &  555.25 & 528.83 & 528.38      & 524.49   \\
s-Trioxane            &  615.72& 609.87   &  595.58& 610.68 & 601.56  &  595.71 & 594.79 & 589.17      & 581.17   \\
Uracil                &  457.42& 453.41   &  440.45& 462.36 & 452.77  &  449.86 & 440.90 & 436.46      & 432.75   \\
Urea                  &  143.82& 142.26   &  140.96& 143.92 & 143.02  &  141.21 & 142.78 & 140.72      & 140.51   \\
\hline
\end{tabular}
}
\end{table}

\begin{table} 
\caption*{\label{tab:sp} Table S10. Lattice energies  (kJ/mol) calculated with QM:QM using single-point computations on the fully-lattice-optimized dispersion-corrected DFT geometries of the X23 molecular crystals.}

\hspace*{-2.5cm}
\scalebox{0.95}{\begin{tabular}{lccccccccc}
\hline
\hline
{\multirow{2}{*}{ X23 structure}}& PBE0:PBE &PBE0   &PBE     & PBE0:PBE & PBE0     & PBE          &B3LYP:BLYP &B3LYP       &BLYP       \\
                      &+D3           &+D3           &+D3     & +MBD     & +MBD     & +MBD         &+D3        &+D3         &+D3        \\
\hline                                                                                                                     
Acetic acid           & 72.8	     &  72.6	    & 72.7   & 73.5	 & 73.5     &  73.6	  &  79.2     &  79.2      &  78.6     \\ 
Adamantane            & 68.9	     &  68.9	    & 69.5   & 71.3	 & 71.4     &  74.2	  &  82.1     &  82.2      &  81.8     \\
Anthracene            & 110.6	     &  110.6	    & 111.0  & 104.8	 & 105.4    &  105.4	  &  126.8    &  126.8     &  126.4    \\
Benzene               & 54.7	     &  54.3	    & 54.7   & 50.9	 & 51.0     &  51.5	  &  64.6     &  64.0      &  64.6     \\
Carbon dioxide        & 24.3	     &  24.1	    & 24.3   & 22.4	 & 22.4     &  22.6	  &  29.7     &  29.7      &  29.3     \\
Cyanamide             & 88.3	     &  88.0	    & 87.7   & 86.2	 & 86.2     &  85.8	  &  94.6     &  94.6      &  94.6     \\
Cyclohexane-1,4-dione & 88.1	     &  87.8	    & 88.2   & 88.8	 & 88.9     &  89.1	  &  101.4    &  101.3     &  100.8    \\
Cytosine              & 159.9	     &  159.7	    & 159.5  & 157.0	 & 157.1    &  156.9	  &  172.8    &  172.8     &  171.1    \\
Ethyl carbamate       & 86.2	     &  86.0	    & 86.2   & 87.6	 & 87.6     &  87.8	  &  94.1     &  94.1      &  93.1     \\
Formamide             & 80.5	     &  80.3	    & 80.2   & 79.5	 & 79.5     &  79.4	  &  86.7     &  86.7      &  86.1     \\
Hexamethylenetetramine& 85.4	     &  85.3	    & 85.4   & 86.2	 & 86.3     &  86.5	  &  96.4     &  96.7      &  94.7     \\
Imidazole             & 89.8	     &  89.6	    & 89.5   & 88.9	 & 88.9     &  88.7	  &  96.8     &  96.8      &  95.8     \\
Naphthalene           & 82.1	     &  82.0	    & 82.3   & 77.4	 & 77.7     &  77.9	  &  94.2     &  94.1      &  93.9     \\
Ammonia               & 38.1	     &  37.9	    & 38.0   & 37.7	 & 37.6     &  37.7	  &  41.5     &  41.5      &  41.6     \\
Oxalic acid ($\alpha$)& 116.2	     &  116.1	    & 116.1  & 117.2	 & 117.2    &  117.2	  &  127.2    &  126.9     &  125.5    \\
Oxalic acid ($\beta$) & 116.0	     &  115.9	    & 116.1  & 116.6	 & 116.5    &  116.5	  &  126.2    &  126.2     &  124.8    \\
Pyrazine              & 63.6	     &  63.4	    & 63.6   & 57.7	 & 57.7     &  57.9	  &  75.8     &  75.8      &  75.3     \\
Pyrazole              & 78.8	     &  78.8	    & 78.9   & 77.9	 & 78.0     &  77.9	  &  85.8     &  85.8      &  84.8     \\
Succinic acid         & 129.1	     &  129.0	    & 129.1  & 131.4	 & 131.4    &  131.3	  &  141.9    &  142.0     &  140.5    \\
s-Triazine            & 59.2	     &  59.1	    & 59.3   & 53.4	 & 53.4     &  53.7	  &  71.5     &  71.5      &  70.8     \\
s-Trioxane            & 55.3	     &  55.1	    & 55.2   & 58.3	 & 58.3     &  58.4	  &  66.6     &  66.7      &  64.9     \\
Uracil                & 137.2	     &  136.9	    & 137.1  & 134.3	 & 134.3    &  134.4	  &  149.3    &  149.2     &  147.9    \\
Urea                  & 111.9	     &  110.8	    & 111.7  & 112.2	 & 111.2    &  112.2	  &  118.0    &  118.0     &  117.2    \\
\hline                                                                                                                                    
RMS /\ mean           &              &  0.3 /\ -0.2 &0.2 /\ -0.02&       &0.3/\ 0.005& 0.7 /\ 0.2 &           &0.2 /\ -0.01 &1.0 /\ -0.8\\
RMS /\ mean (\%)      &              &  0.4 /\ -0.3 &0.3 /\ -0.02&       &0.2 /\ 0.06& 0.9 /\ 0.4 &           &0.2 /\ -0.02 &1.0 /\ -0.8\\
\hline
\hline
\end{tabular}}
\end{table} 

\begin{figure}[!]
\begin{center}
\caption*{\label{pbed3} Figure S1. Histograms showing an error distribution of {\it{X23}} cell volumes  with respect to experiment (without thermal correction) using  (a) DFT+D3 and PBE0:PBE(PBE0+D3) (b) DFT+MBD and PBE0:PBE(PBE0+MBD) (c) DFT+D3 and B3LYP:BLYP(B3LYP+D3)}
\hspace{-1.5cm}
\scalebox{0.65}{\includegraphics[angle=0,origin=c]{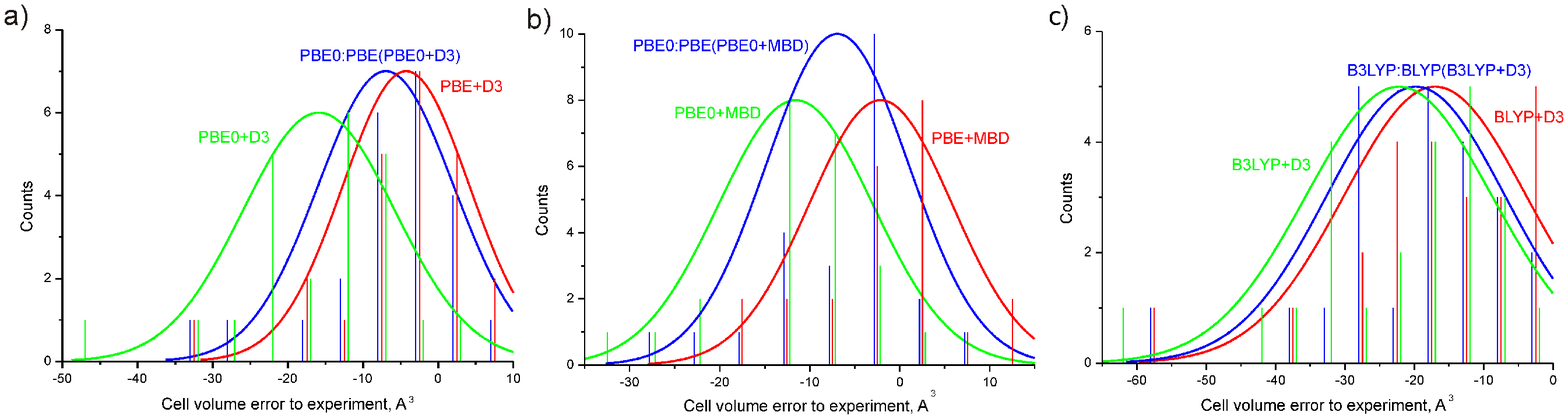}}
\end{center}
\end{figure}

\clearpage
\bibliographystyle{apsrev}